\documentclass[useAMS,usenatbib]{mnras}
\usepackage{graphicx}
\usepackage{graphics}
\usepackage{multicol}
\usepackage{etex}
\usepackage{setspace}
\usepackage{float}
\usepackage{pdflscape}
\usepackage{multirow}
\usepackage{hhline,booktabs,amsmath}
\usepackage{makecell}
\usepackage{caption}
\usepackage{subcaption}
\usepackage{newtxtext,newtxmath}
\usepackage{ae,aecompl}

\usepackage{graphicx}
\usepackage{rotating}
\usepackage{epsfig}
\usepackage{amssymb}
\usepackage{amsmath}
\usepackage[utf8]{inputenc}
\usepackage{array}
\usepackage{ragged2e}
\usepackage{longtable}
\usepackage{supertabular}
\usepackage{subfloat}
\usepackage{graphicx}	% Including figure files
\usepackage{amsmath}	% Advanced maths commands
\usepackage{amssymb}	% Extra maths symbols

\title{Coronal Vertical Structure Variations in Normal Branch of GX 17+2: AstroSat's SXT and LAXPC perspective}

\author[S. Malu et al.]{
Malu S,$^{1}$\thanks{E-mail:malu.sudhaj@gmail.com}
K. Sriram,$^{1}$
V. K. Agrawal,$^{2}$
\\
$^{1}$Department of Astronomy, University College of Science, Osmania University, Hyderabad, India\\
$^{2}$Space Astronomy Group, ISITE Campus, U R Rao Satellite Center, 560037, Bangalore, India \\
}

\date{Accepted XXX. Received YYY; in original form ZZZ}

\pubyear{2020}

\begin{document}
\label{firstpage}
\pagerange{\pageref{firstpage}--\pageref{lastpage}}
\maketitle

% Abstract of the paper
\begin{abstract}
We performed spectro-temporal analysis in the 0.8--50 keV energy band of the neutron star Z source GX 17+2 using {\it AstroSat}
 Soft X-ray Telescope (SXT) and Large Area X-ray Proportional Counter (LAXPC) data.
The source was found to vary in the normal branch of the Hardness Intensity Diagram. Cross-correlation studies of LAXPC light curves
 in  soft and hard X-ray band
unveiled anticorrelated lags of the order of few hundred seconds.
For the first time, Cross-correlation studies were performed using SXT soft and LAXPC hard lightcurves and they
 exhibited correlated and anti-correlated lags of
the order of a hundred seconds.
Power density spectrum displayed NB oscillations of 6.7--7.8 Hz (quality factor 1.5--4.0).
 Spectral modeling resulted in inner disk radius of $\sim$ 12--16 km  with $\Gamma$ $\sim$ 2.31--2.44 indicating that disk is close to the ISCO and a similar value of 
 disk radius was noticed based on the reflection model.
Different methods were used to constrain the corona size in GX 17+2. Using the detected lags, corona size was found
to be 27-46 km ($\beta$ = 0.1, $\beta$ = v$_{corona}$/v$_{disk}$) and 138--231 km ($\beta$ = 0.5). Assuming
the X-ray emission to be arising from the Boundary Layer (BL), its size was determined to be 57--71 km. Assuming that BL is ionizing 
the disk's inner region, it's size was constrained to $\sim$ 19--86 km. Using NBO frequency,
the transition shell radius was found to be around 32 km.
Observed lags and no movement of the inner disk front strongly indicates that the varying corona structure is causing the X-ray variation in the
 NB of Z source GX 17+2.
\end{abstract}

\begin{keywords}
accretion, accretion disk---binaries: close---stars: individual (GX 17+2)---X-rays: binaries
\end{keywords}

%%%%%%%%%%%%%%%%%%%%%%%%%%%%%%%%%%%%%%%%%%%%%%%%%%

%%%%%%%%%%%%%%%%% BODY OF PAPER %%%%%%%%%%%%%%%%%%

\section{Introduction}
Z sources are highly luminous Neutron Star (NS) Low Mass X-Ray Binaries (LMXBs) with their luminosities approaching Eddington Luminosity or exceeding it. They trace out a Z track in the 
Hardness Intensity Diagram (HID)/ colour-colour Diagram (CCD) (Hasinger \& Van Der Klis 1989).
There are three main branches to this track viz. the horizontal, normal and flaring branch (HB, NB and FB) with a hard apex at the horizontal to normal branch transition and
a soft apex at the normal to flaring branch transition (Hasinger \& van der Klis 1989; van der Klis 2006). Z sources are further grouped into two main categories - Sco X-1 like and Cyg X-2 like sources. This classification
is based on the characteristics of the Z track traced out by them (Kuulkers et al. 1994, 1997). 
Earlier multiwavelength studies indicated that $\dot{m}$ increases from  HB to FB via NB (Hasinger et al. 1990; Vrtilek et al. 1990) and later Homan et al. (2002) argued that $\dot{m}$ might be constant along different branches on the Z track. A contrary picture was given by Church et al. (2008) arguing that $\dot{m}$ increases from FB to HB through NB. Based on the spectral fits of XTE J1701-462, Lin et al. (2009) concluded that $\dot{m}$ is constant along the different branches however each one is associated with different physical mechanisms. This study clearly showed that $\dot{m}$ decrease as 
the source traverses from Cyg-like to Sco-like followed by atoll sources.

Spectral modelling of Z sources have been done through several models all of which seem to fit the spectra well but have starkly different Accretion Disk Corona (ADC) geometry. The origin of soft and
hard X-ray photons has been a matter of debate and several models have been proposed like the eastern, western and hybrid models.
The eastern model uses a multi-temperature black body emission (MCD) from the disk which supply seed photons for inverse comptonization in a small coronal region/ hot electron cloud in the inner region of the accretion disk
(Mitsuda et al. 1984). The western model (White et al. 1986) uses a single temperature black body emission from the neutron star (NS) surface or close to the surface and a
high energy power law component radiated by the comptonization occurring in the inner most region of the accretion disk.
Church \& Bałucinska-Church (1995, 2004) proposed a model incorporating a black body emission from the NS surface or an optically thick boundary layer and a comptonization of soft photons from the disk
 by hot electrons in an extended corona above the disk. 
The hard tail $>$ 30 keV in the spectrum of these sources has also been interpreted in the framework
of the bulk motion comptonization model (Farinelli et al. 2008). 
The jet model on the other hand suggests that the hard component could be originating from the comptonization of
soft seed photons by the high energy electrons of a jet (Di Salvo et al. 2002, Reig \& Kylafis 2016). Radio jets are often found to be present when the source is in the horizontal branch and the upper normal branch (eg. Migliari et al. 2007; 
Di Salvo et al. 2002) Popham \& Sunyaev (2001) propose a model where the comptonized spectrum arises from a hot, low density boundary layer around the NS surface and the black body spectrum arises from an optically thick accretion disk further supported by the Fourier frequency resolved spectrum (Gilfanov et al. 2003; Revnivtsev \& Gilfanov 2006). 
 Based on the spectral analysis of Aql X-1 and 4U 1608-52, Lin et al. (2007)
used a hybrid model which uses a single temperature blackbody (BB) plus a broken power law (BPL) for the hard state, and two thermal components (MCD and BB) along with a constrained BPL for the soft state.
Although several models have been attempted so far, the accretion disk-corona geometry of NS LMXB is still an open question including the structure of corona or the boundary layer.

Each branch in the Z track is associated with a characteristic Quasi Periodic Oscillation (QPO, Hasinger \& van der Klis 1989).  
The horizontal branch oscillations (HBOs) vary in the range of 15-60 Hz, whereas normal branch oscillations (NBOs) vary in a very limited range of 5-8 Hz (van der Klis 2006). So far, only two sources, namely, Sco X-1 and GX 17+2 
have exhibited flaring branch oscillations (FBOs)  (Sco X-1: Priedhorsky et al. 1986; GX 17+2: Penninx et al. 1990, Homan et al. 2002). FBOs occur in a small region of the flaring branch closest to the 
soft apex with a frequency ranging from 10--25 Hz. Kilo-Hertz QPOs (200-1200 Hz) are also found to be associated with Z sources such as GX 17+2 (Wijnands et al. 1997), Sco X-1 (van der Klis et al. 1996), 
Cyg X-2, GX 5-1 (Wijnands et al. 1998a,b), GX 340+0 (Jonker et al. 1998), GX 349+2 (Zhang, Strohmayer \& Swank 1998) and XTE J1701-462 (Sanna et al. 2010). 

HBOs have been considered to be the beat frequency of the keplerian orbital frequency of the inner edge of the disk and the spin frequency of the neutron star (Alpar \& Shaham 1985, Lamb et al. 1985). %But,
The Relativistic Precession Model proposes that HBOs are the nodal precession of tilted orbits near the Neutron Star (NS) 
(Stella \& Vietri 1998a, 1999). The model suggests that QPOs are basically the fundamental frequencies of a test particle/blob moving near the NS. It also explains the kHz QPOs as the relativistic 
periastron precession of eccentric orbit. Models for the appearance of NBOs/FBOs have been fairly ambiguous with early models associating them to oscillations in the optical depth of accretion flow in the inner disk region 
(Lamb 1989, Fortner, Lamb \& Miller 1989) or the oscillations of sound waves in a thick disk (Alpar et al. 1992). Both models require near Eddington luminosities to work and therefore have fallen short of 
explaining the NBO/FBO like QPOs seen in an atoll source like 4U 1820-30 (Wijnands, van der Klis, \& Rijkhorst 1999) where luminosities are less than half the Eddington luminosity (Homan et al. 2001). 
 Moreover NBOs are smoothly transformed into FBOs and hence thought to have
common origin (Kuulkers et al. 1997; Casella et al. 2006) but this warrants further investigation of the
spectral and timing properties of these sources.

Titarchuk et al. (2001) explains NBOs as acoustic oscillations of a spherical shell around the neutron star (NS) surface which forms when the disk is destroyed due to radiation pressure near the Eddington accretion rate.
Motta \& Fender (2018) found that the ejection of the ultra-relativistic flow were associated with the simultaneous appearance of NBO ($\sim$ 6 Hz) and HBO ($\sim$ 43 Hz) in the X-ray power density spectrum.
 NBO strength has been associated with the inclination of the system (Kuulkers \& van der Klis 1995). Kuulkers \& van der Klis suggested that lower inclination sources show stronger NBOs. 
 Also, the observation of FBOs in GX 17+2, with an inclination less than 45$^{\circ}$, is in agreement with this (Kuulkers et al. 1997).

One of the important timing tools to constrain inner accretion disk geometry in X-ray binaries is Cross Correlation function (CCF) study between soft and hard X-ray energy bands (eg. Vaughan et al. 1999; Sriram et al. 2007; Sriram et al. 2012).
 Milli second lags in CCFs are considered to be the comptonization timescales of soft photons up scattered to hard photons (Vaughan et al. 1999; 
Kotov et al. 2001; Reig \& Kylafis 2016). However lower Fourier frequency CCF studies carried out in Z sources have associated few hundred seconds lags to the HB/NB branches of the Z track
(Cyg X-2: Lei et al. (2008), GX 5-1: Sriram et al. 2012, GX 17+2: Sriram et al. 2019). The detection of few hundred seconds lags in HB and NB are often seen to be associated with the presence of a compact corona or a jet but a clear picture is lacking.
(Migliari et al. 2007, 2011; Fender et al. 2009; Sriram et al. 2019). 
Based on the CCF studies, Sriram et al. (2019) proposed a model where the few hundred second lags found in such sources could primarily be 
the readjustment time scales of the corona/non keplerian flows.

 The source is located at a distance of $\sim$13 kpc (Galloway et al. 2008) and is a Sco X-1 like NS LMXB Z source with a spin frequency of 293.2 Hz (Wijnands et al. 1997). Its optical counterpart is yet
to be unambiguously confirmed. GX 17+2 is a low inclination system with i $<$ 40$^{\circ}$ (Cackett et al. 2010, Ludlam et al. 2017a). Although GX 17+2 has been thoroughly studied previously,
the soft part of the spectrum ($<$ 3 keV) remains less explored. Bepposax spectrum of GX 17+2 in the energy range 0.1-200 kev was found to exhibit a hard tail at energies above $\sim$ 30 keV  when the source was in the horizontal branch whereas the hard tail was undetectable
when the source moved towards the NB (Di Salvo et al. 2000). A power law index of 2.7 was found to adequately model the tail and an effective black body radius of $\sim$ 40 km was found. 
{\it Chandra} HEG spectra was modelled by Cackett et al (2009) using a Diskbb+bbody model in the 3-9 keV energy range along with an iron line.

For Suzaku observations of GX 17+2 (Cackett et al. 2010) phenomenological modeling was performed using the relativistic diskline model to constrain the inner disc radius to 7-8 GM/c$^2$ and inclination to 15$^{\circ}$-27$^{\circ}$.
Ludlam et al. (2017a) modeled the NuStar spectra of GX 17+2 in the energy range 3-30 keV and found that the inner disk radius extends upto 1.00-1.02 ISCO (for spin 0) and the
source inclination was found to be 25$^\circ$-38$^\circ$. Agrawal et al. (2020) found that the power law component follows a decreasing trend as the source moves from HB to NB and later increases as it
moves from NB to FB. They also found that the hard X-Ray tail ($>$ 30 keV) contributes to 17 \% of the flux in the upper HB while in the NB this fraction is just 3-4 \%. 
The black body radius estimated from their spectral model is $\sim$ 8-10 km, which is very close the NS radius and inner disk radii was to be 28-42 km.

Since soft energy band (< 3 keV) is essential for both spectral and timing properties of Z sources, we report the simultaneous spectral and timing studies of GX 17+2 using {\it SXT} and {\it LAXPC} of the {\it AstroSat}. Here we report the CCF among SXT and LAXPC energy bands, PDS and broadband spectrum (0.8-50 keV) and discuss the results to explain the accretion disk-corona geometry of GX 17+2 in the normal branch.

\section{Data Reduction and Analysis}

Regular pointing observations of GX 17+2 were performed using AstroSat for 11 satellite orbits from August 22, 2017 23:19:44.20 to August 24, 2017 01:18:36.66
(Observation ID AO3\_072T01\_9000001484) for an effective exposure time of $\sim$ 32 ks. The data was collected during
AO CYCLE 3. Observations were performed using LAXPC and SXT instruments of the AstroSat . The LAXPC is the  Large Area X-ray Proportional Counter (LAXPC) which consists of 3 identical proportional counter units (LAXPC 10,20 and 30) having a total effective area $\sim$ 6000
cm$^2$ at 15 keV operating in the energy range 3-80 keV (Yadav et al. 2016a; Agrawal et al. 2017; Antia et al. 2017). It has a timing resolution of 10 $\mu$s and a dead time of 42 $\mu$s.
The SXT is the Soft X-ray telescope operating in the 0.3-8 keV energy range with an energy resolution of $\sim$ 150 eV and an effective area of $\sim$ 128 cm$^2$ at 1.5 keV (Singh
et al. 2017). The Photon Counting (PC) mode of SXT has a time resolution of 2.4 s and the Fast Window (FW) mode has a time resolution of 0.278 s.

LAXPC software provided by  AstroSat Science Support Center (ASSC) was used for reducing the obtained LAXPC level 1 data in Event analysis (EA) mode.
Event files, filter files and GTI files (with Earth occulation and SAA removed) were produced following standard procedures as instructed in the 
LAXPC software (Format A, May 19, 2018 version)\footnote{http://astrosat-ssc.iucaa.in/?q=laxpcData}, with the 
laxpc\_make\_event routine being used for generating event file and laxpc\_make\_stdgti routine used for generating the good time intervals (GTI). These were later used 
for producing light curves and spectra using the routines laxpc\_make\_lightcurve and laxpc\_make\_spectra. The software also generates the necessary response files for each of the LAXPC units.
SXT level 2 data in the PC mode was processed by initially using the event merger code to produce a merged cleaned event file and the sxtARFmodule to produce appropriate ancillary response files.
XSELECT was used for then producing the image, light curves and spectra. Response file (sxt\_pc\_mat\_g0to12.rmf) and deep blank sky background spectrum
file (SkyBkg\_comb\_EL3p5\_Cl\_Rd16p0\_v01.pha) provided by the SXT team was used during spectral analysis \footnote{https://www.tifr.res.in/$\sim$astrosat\_sxt/dataanalysis.html}. 
As count rate was found to be $>$ 40 cts s$^{-1}$, source was extracted using a 3$\arcmin$-10$\arcmin$ annulus in order to account for the pile up effect (see AstroSat handbook \footnote{http://www.iucaa.in/$\sim$ astrosat/AstroSat\_handbook.pdf}) and
background light curve was extracted using a 13$\arcmin$-15$\arcmin$ annulus (see Figure 1). LAXPC 10 and 20 data has been used for timing analysis and 
LAXPC 10 alone for spectral analysis which is well calibrated and has less background issues (Agrawal et al. 2020). LAXPC 30 was not used for analysis
owing to a gas leakage in LAXPC 30 which leads to response uncertainties (Antia et al. 2017).
LAXPC 10 top layer spectra alone was used for spectral analysis in order to minimize the background (eg. Beri et al. 2019).  
LAXPC spectrum in the range 4-50 keV and SXT data in the range 0.8-7 keV was used for the joint spectral analysis. Data below 0.8 keV were not considered
due to uncertainties in response (Bhargava et al. 2019). A systematic error of 3\% was added while performing spectral fitting \footnote{https://www.tifr.res.in/$\sim$astrosat\_sxt/dataana \_up/readme\_sxt\_arf\_data\_analysis.txt} (eg. Jithesh et al. 2019, Bhargava et al. 2019).

\section{Timing analysis}

Combined light curves of GX 17+2 from LAXPC 10 and 20 were extracted for performing timing analysis (Figure 2). HID was obtained using the hard colour 10.5-19.7 keV/7.3-10.5 keV and intensity in 
7.3-19.7 keV energy range (Homan et al. 2002). During this observation, HID reveals only the Normal Branch (Figure 3) (in agreement with Agrawal et al. 2020). It has been divided into upper (A), middle (B) and lower (C) portions to study the evolution of the source along the track.

Cross Correlation Function (CCF) studies were performed between soft and hard X-ray light curves from LAXPC and SXT. The {\it crosscor} tool in the XRONOS package was 
used for the CCF analysis (see Sriram et al. 2007, 2011a; Lei et al. 2008). 

CCF lags were obtained between 3 - 5 keV and 16-20 keV, 3-5 keV and
20-40 keV and 3 - 5 keV and 20-50 keV lightcurves using a 30 second bin size.
Hard lag means that the hard photons are lagging to soft photons and vice-versa for soft lags.
Similar analysis were performed between the 0.8-2 keV of SXT vs. 10-20 keV, 16-20 keV, 20-40 keV and 20-50 keV of LAXPC light curves. 
Simultaneous SXT and LAXPC light curve
segments were used to study CCF. CCFs were fitted using a Gaussian function to obtain the lags and errors were estimated
with 90\% confidence level using the the criterion of $\Delta$$\chi$$^{2}$ = 2.7.

The 3-5 keV vs. 16-20 keV, 20-40 keV and 20-50 keV light curves show anti-correlated CCFs
in three separate segments among which two of them show hard lags of 
139 $\pm$ 31 s and 217 $\pm$ 20 s (3-5 vs 16-20 keV) with cross-correlation coefficients (CC) of -0.57 $\pm$ 0.25 and -0.64 $\pm$ 0.11 respectively 
(Figure 4 and Table 1a). Remaining segments are positively correlated with a CC ranging from
0.4-0.8 with no lags. 

The 0.8-2 keV SXT vs. 10-20 keV, 16-20 keV, 20-40 keV and 20-50 keV LAXPC light curves exhibited lags on few occasions but the errors were quite large for any 
confirmation (eg. 0.41 $\pm$ 0.32), 
except in one segment where an anti-correlated soft lag of -133 $\pm$ 34 s (0.8-2 vs 10-20 keV) with a CC of -0.41 $\pm$ 0.19 (Table 1b) and a correlated hard lag of 69 $\pm$ 40 s 
with a CC of 0.33 $\pm$ 0.17 was noted in the  0.8-2 vs. 10-20 keV CCF (figure 5), although in the higher energy bands (16-20 keV, 20-40 keV and 20-50 keV 
) this feature became statistically insignificant with higher error bars (Lag $\sim$ 53 $\pm$ 48 s, 41 $\pm$ 20 s, 35 $\pm$ 26 s respectively).
In most of the sections, CCFs were uncorrelated with CC varying from 0.35$\pm$0.22.

Power Density Spectrum (PDS) was obtained in the 3-20 keV energy band to search for characteristic QPOs in the obtained portions. 
Power spectra were normalized in units of (rms/mean)$^{2}$/Hz (Miyamoto et al. 1991).
 
The final 3 segments of the light curve (Figure 2) which are associated with the lower most portion of the Normal Branch (Figure 3), show signatures of a $\sim$ 7 Hz NBO. 
The PDS was fitted using power-law+Lorentzian model. The first of the these three exhibit a broad feature at 7.68 $\pm$ 0.35 Hz with a Q factor $\sim$ 1.95, second of 
the last three show a similar feature at 7.75 $\pm$ 0.47 Hz with a Q factor $\sim$ 1.42 (Figure 6). The last section shows a prominent NBO at a frequency of 6.88 $\pm$ 0.21 Hz with a 
Q factor $\sim$ 3.95 (Figure 6c). A similar NBO value of 7.42 $\pm$ 0.23 Hz was found by Agrawal et al. (2020) using AstroSat LAXPC data of GX 17+2.

Following which a dynamical power density spectrum was obtained (Figure 7) with a 40 s time resolution and 0.25 Hz frequency resolution, in order to verify the appearance of the NBOs along the time
series evolution. Dynamic PDS clearly shows the
the appearance of the $\sim$ 7 Hz NBO in the last $\sim$ 3500s. The presence of a low frequency noise component around 1-5 Hz is also noticed. 

\section{Spectral analysis}   
Spectral analysis was performed for each of the three sections (A-C, upper to lower) shown in HID using XSPEC v12.10.0c (Arnaud 1996). 
Simultaneous SXT+LAXPC joint spectral fitting in the energy range 
0.8-50 keV was performed. Figure 9 gives the SXT + LAXPC joint spectral fit, where the top panel gives the unfolded spectra (thick line) along with the model
components (dashed line) and the bottom panel gives the residuals to the fit.

Absorption column density N$_H$ was fixed at $\sim$ 2 $\times$ 10$^{22}$ (Di Salvo et al. 2000) using the {\it Tbabs} model (Wilms et al. 2000)  and a constant factor was involved in order to 
account for the relative normalization between SXT and LAXPC data. The gain fit command was used to account for the gain correction in the SXT spectrum.
While performing gain fit, slope was fixed at unity and the offset was allowed to vary as suggested by the SXT instrument team \footnote{www.tifr.res.in/$\sim$astrosat\_sxt/dataana \_up/readme\_sxt\_arf\_data\_analysis.txt}.

Cackett et al. (2008) based on the {\it Chandra} and {\it RXTE} data of few Z sources found that the continuum was adequately described by a 
a blackbody ({\it bbody}) +  disk blackbody ({\it Diskbb})+  powerlaw component during the soft/intermediate state and a {\it bbody} + broken-powerlaw model
 during the hard state. They followed the prescription of Lin et al. (2007), who based on their investigation of the luminosity dependence of 
temperatures measured for different continuum models (for atoll sources),
suggested using a similar model. Cackett et al. (2010) later replaced the broken power-law with a simple power-law model for the spectra of
 GX 17+2 and few other Z sources obtained using {\it Suzaku} and {\it XMM-Newton}.

Initially we modeled the continuum using a {\it Diskbb+power-law} model.
Residuals from the fit clearly indicated an emission feature at $\sim$ 6.7 keV which was fitted using a Gaussian model whose centroid energy 
was fixed at 6.7 keV. 

 Apart from which a residual at $\sim$ 32-33 keV was noted in Section A and B, which could be the Xenon K emission feature of 
instrumental origin  (Antia et al. 2017).
We fitted a Gaussian component to account for it (eg. Sridhar et al. 2019, Sharma et al. 2020). 
We found the central line energy to be $\sim$ 32.4 keV and $\sigma$ to be $\sim$ 1.5 keV. This Gaussian component
parameters were fixed while fitting the other model parameters. This model resulted in a high $\chi$$^2$/dof value of 1086/790 (First section).
Upon which we added a single temperature {\it bbody} model which resulted in a significantly reduced $\chi$$^2$/dof value $\sim$ 995/788.
Similar reduction in reduced $\chi$$^2$ values were noted in the other 2 sections too  $\sim$ 1017/788 and 1071/788 respectively for B \& C sections (see Table 2). 
The power-law index $\Gamma_{PL}$ varied from 2.31$_{-0.03}^{+0.03}$ to 2.46$_{-0.03}^{+0.03}$ from A to C. 
Hence the overall model used was {\it Diskbb + bbody + Gaussian (Fe) + Gaussian (Xe K)+ power-law} (Table 2).
The equivalent width of the iron line was found to be 71 eV $_{-50}^{+46}$, 83 eV $_{-51}^{+50}$ and 117 eV $_{-54}^{+47}$ 
for the A,B and C sections respectively.

We then used {\it bbrefl} (Ballantyne 2004) to model the reflection from the disk by a blackbody (see table 3) and it was relativistically 
blurred with the convolution model {\it rdblur} (Fabian et al. 1989) (eg. Mondal et al. 2016, Cackett 2016).  
Emissivity index was fixed at -3, outer disc radius at 1000 Rg (Rg=GM/c$^{2}$) and the inclination angle was allowed to vary.
We constrained the reflection fraction f$_{refl}$ to 0.62--0.66, the ionization parameter log$\xi$ to 3.04--3.37 and 
the incident blackbody temperature $\sim$ 2.37--2.45 keV. 

Disk inclination angle and inner radius from the {\it rdblur} model was constrained
by using the steppar command in xspec which computes the $\Delta$$\chi$$^2$ (=$\chi$$^2$-$\chi$$^2$$_{min}$) for each of the parameters.
Figure 8a,b shows $\Delta$$\chi$$^2$ vs inclination angle and inner radius R$_{in}$ ({\it rdblur}) for the best fit model (Table 3). 

Based on the {\it rdblur} model, R$_{in}$ was found to be 7.51--8.06 Rg (15.6 -- 16.7 km). 
The iron abundance of 1 was used in the model. Therefore, the overall best fit model used was {\it Diskbb + rdblur*bbrefl + Gaussian (Xe K)+ power-law}.
This led to $\chi$$^{2}$/dof values of 957/787, 977/787 and 1044/787 for A,B and C sections respectively.

We also tried the {\it Nthcomp + Power-law} model (Zdziarski et al. 1999, Zycki et al. 1999) based on Agrawal et al. (2020) and the
 results are shown in Table 4. The Compton cloud /corona temperature is found to be around 2.6 keV, asymptotic power-law index of nthcomp was found to be varying 
from 2.19 to 2.69 along with the steepening of the power-law index from 2.56 to 2.72. 
A blackbody component was added to the above model but it did not lead to any improvements in the fit.

\section{Results and Discussion}

\subsection{CCF lags and Coronal size Determination}
Simultaneous spectra for GX 17+2 in the soft (<7 keV) and hard (>7 keV) energy bands has not been explored substantially before, 
which we achieve with the SXT (0.8--7 keV) and LAXPC (4--50 keV) data. Although satellites have covered this energy domain,
 CCF studies have never been done before.
For the first time we report the CCF of GX 17+2 between 0.8--2.0 keV (SXT) vs 10-20 keV, 16-20 keV, 20-40 keV and 20-50 keV. A
-133 $\pm$ 34 s anti-correlated soft lag and 69 $\pm$ 40 s (0.8 vs 10-20 keV) correlated hard lag was found in one of the segments using the simultaneous 
SXT soft and LAXPC hard light curve (see Table 1).
Using LAXPC soft and hard light curves, anti-correlated hard lags were found in two separate light curve 
segments with lag values of 139 $\pm$ 31 s and 217 $\pm$ 20 s (CC $\sim$ -0.6) (3-5 vs 16-20 keV).

Similar lags were reported in other Z sources using RXTE and NUSTAR in 3-5 vs 16-20 keV energy band (Lei et al. 2008; Sriram et al. 2019; 2012).
The hard component is most probably arising from the compact corona lying in the inner region of the
 accretion disk or the boundary layer over the neutron star surface (Popham \& Sunyaev, 2001).
The lags are interpreted as the readjustment time scale of the compact corona which is varying in size. 

The detected lags of few hundred seconds in CCF cannot be just the readjustment time scale of radial and vertical readjustment structure 
of the Keplerian portion of the disk which is of the order few tens of seconds and hence Sriram et al. (2019) concluded that 
the readjustment time scale of the coronal structure also needs to be considered. 
Moreover our spectral results suggest that the inner disk front is not moving and is located at $\sim$ 12-16 km (from Diskbb normalization 
and for i $\sim$ 27$^\circ$).
during the NB. Observed hard lags are therefore due to the variation in size of the corona. A similar picture of varying corona structure 
without any movement of the disk was discussed by Kara et al. (2019) in case of a BH source 
MAXI J1820+070.
Under the assumption that the readjustment velocity in the coronal region v$_{corona}$=$\beta$v$_{disk}$ and $\beta$ is
$\le$ 1 as the coronal viscosity is less than the disk viscosity, the below equation for coronal 
height was derived, which considers the lag to be a combination of the disk and coronal readjustment. 
\begin{equation}
H_{corona}=\Bigg[\frac{t_{lag} \dot{m}}{2 \pi R_{disk} H_{disk} \rho}-R_{disk}\Bigg] \times \beta \; cm
\end{equation}
where H$_{disk}$ = 10$^{8}$ $\alpha^{-1/10}$ $\dot{m}_{16}^{3/20} R_{10}^{9/8} f^{3/20} $ cm, 
$\rho$ = 7 $\times$ 10$^{-8}$ $\alpha^{-7/10}$ $\dot{m}^{11/20}$ $R^{-15/8}$ $f^{11/20}$  g cm$^{-3}$, f = (1-(R$_s$/R)$^{1/2}$)$^{1/4}$ 
 (Shakura \& Sunyaev 1973, Sriram et al. 2019).

The coronal height was estimated for the segments of the light curve which exhibited CCF lags. Here the disk 
radius R$_{disk}$ was considered to be 16 km based on the radius obtained from the spectral model (see Table 3)  
and $\beta$ was taken to be 0.1--0.5 (Manmoto et al. 1997, Pen et al. 2003, McKinney et al. 2012). The coronal height was thus 
estimated to be $\sim$ 27--46 km ($\beta$ = 0.1) and 138--231 km ($\beta$ = 0.5).

\subsection{Inner disk, Spherization and Boundary layer Radii}

The Diskbb model normalization (Mitsuda et al. 1984) is proportional to the inner disk radius via the relation 
R$_{in}$ (km) = $\sqrt{(N / cos i)}$ $\times$ D / 10 kpc. 

The estimated radii are not true radii and need to be corrected for spectral hardening factor ($\kappa$ $\sim$ 1.7--2.0; Shimura \& Takahara 1995) 
along with inner boundary condition correction factor($\xi$ = 0.41; Kubota et al. 1998) which enhances the true radii by a factor of 1.18--1.64 
(R$_{eff}$ = $\kappa^{2}$ $\xi$ R$_{in}$; Kubota et al. 2001). 
We estimate R$_{eff}$ to be around $\sim$ 9-14 km (for i = 25$^\circ$, for correction of 1.18--1.64) for A, B \& C sections 
respectively (Table 2) and using the same relation we obtain R$_{eff}$ $\sim$ 12--16 km for all the sections respectively for  
similar correction factors (Table 3). 

Nustar study of GX 17+2 shows a kT$_{in}$ $\sim$ 1.92 keV similar to the value seen in our 
observations (see Table 2) but with a slight difference in disk normalization values (Ludlam et al. 2017). Suzaku spectra 
of GX 17+2 observed the disk with a temperature kT$_{in}$ $\sim$ 1.75 keV along with a normalization N$_{Diskbb}$ $\sim$ 86 and found a dissimilarity 
in the true disk radius from Diskbb model and radius arrived from reflection model (Cackett et al. 2010). Based on Chandra + RXTE 
spectral fits, Cackett et al. (2009) observed kT$_{in}$ $\sim$ 1.57 keV and N$_{Diskbb}$ $\sim$ 150. Ferinelli et al. (2005) reported
a blackbody radius of about $\sim$ 35 km for GX 17+2 in the normal branch. 
 
With rdblur model, we found R$_{in}$ to be at 
$\sim$ 16 km ($\sim$8 R$_{g}$) (Table 3) and this value does not vary much if we change the other parameters of this model. 
This value is close to the value of R$_{eff}$ from {\it Diskbb} model indicating that the disk is close to the ISCO which is often the case for 
GX 17+2 (Ludlam et al. 2017; Cackett et al. 2010).

Since the source was found to emit 1.1--1.3 L$_{Edd}$ based on our spectral analysis 
(see Table 3, considering L$_{Edd}$ = 3.80 $\times$ 10$^{38}$ ergs s$^{-1}$ (Kuulkers et al. 2003)), 
the inner region of the disk would be puffed up due to radiation pressure (Shakura \& Sunyaev, 1973). Therefore the radius of this
quasi spherical region of the radiation-pressure dominated disk was estimated using the relation R$_{sp}$ = 32 $\dot{m}$$_{Edd}$ km (Ding et al. 2011).
Based on the $\dot{m}$ determined from luminosities obtained from spectral analysis (L = GM$\dot{m}$/R), 
R$_{sp}$ is found to be $\sim$ 36--43 km.

Boundary layer (BL) is the region between the fast rotating accretion disk and relatively slow spinning neutron star surface (Popham \& Sunyaev, 2001). 
This layer could be responsible for the hot blackbody or the comptonized component that dominates around 7--20 keV (e.g. Popham \& Sunyaev 2001, Barret et al. 2000).
The emission from this boundary layer could significantly contribute to the accretion luminosity. Popham \& Sunyaev (2001) model proposes a boundary layer structure whose
radius and height varies with mass accretion rate as given by the equation,

\begin{equation}
log(R_{BL} - R_{NS}) \sim 5.02 + 0.245 \Bigg[log\Big({\frac{\dot{M}} {10^{-9.85}\ \ M_{\odot} yr^{-1}}\Big)}\Bigg]^{2.19}
\end{equation}

Based on this equation we estimated the BL radius to be 57--71 km for the C--A sections, using $\dot{M}$ obtained from 
luminosity in the 0.8--50 keV energy range (eg. Sriram et al. 2019, Ludlam et al. 2017b), using the equation
 L= $\frac{GM\dot{M}}{R}$ with M = 1.4 M$_{\odot}$ and R = 10 km.

As per the standard disk theory by Shakura \& Sunyaev (1973), the BL luminosity should be related to the disk luminosity as (Pringle 1981),

\begin{equation}
L_{BL}=(1-\beta^{2})\frac{GM\dot{M}}{2R_{NS}}
\end{equation}

here $\beta$ is the NS angular velocity in Keplerian units ($\Omega_{*}$/$\Omega_{k}(R_{NS})$), G is the gravitational constant, 
M is the mass of the NS, R$_{NS}$ is the NS radius and $\dot{M}$ is the mass accretion rate. 
But Kluzniak (1987) showed that part of the BL's kinetic energy goes into the spinning of the star and therefore
correct L$_{BL}$ should be,

\begin{equation}
L_{BL}=(1-\beta)^{2}\frac{GM\dot{M}}{2R_{NS}}
\end{equation}

Using the spin frequency of GX 17+2 (293.2 Hz; Wijnands et al. 1997), the angular velocity was estimated which resulted in $\beta$ $\sim$ 0.13.
R$_{NS}$ value of 10 km was used for the calculation. By initially considering the BB+PL luminosity obtained from spectral fitting as L$_{BL}$ (assuming the BL to be responsible for both the
 hot BB and comptonized component e.g. Popham \& Sunyaev 2001, Barret et al. 2000), we estimated the
mass accretion rate $\sim$ 2.36--3.42 $\times$ 10 $^{18}$ g/s and determined the radius of the BL from equation 2 as 62--104 km. Later by considering only the hot BB component to be arising from the boundary layer, the BB luminosity obtained from spectral fitting was substituted 
as L$_{BL}$ to estimate the mass accretion rate $\sim$ 1.23--2.10 $\times$ 10 $^{18}$ g/s from equation 4. 
This yielded the radius of the BL from equation 2 as 30--54 km.

Using the luminosity (L$_{BL}$) obtained from hot BB flux, height (Z) of the ionizing source (BL) above the disk was estimated using the equation 
derived by Cackett et al. (2010) as,

\begin{equation}
Z^{2} = \frac{L_{BL}}{n\xi} - R_{in}^{2}
\end{equation}

In order to constrain Z, n $>$ 10$^{21-22}$ cm$^{-3}$ (Shakura \& Sunyaev, 1973)
and log$\xi$ obtained from our spectral fitting (see Table 3)  were used. 
Based on R$_{in}$ = 16 km, Z was constrained to be 19.3 km, 19.5 km, 23.1 km 
(for n = 10$^{22}$ cm$^{-3}$) and 77.5 km, 76.1 km, 86.6 km (for n = 10$^{21}$ cm$^{-3}$ ) for the three sections from A to C.

\subsection{NBO detection in the PDS}

A spherical structure around the NS surface is suggested by Hasinger (1987) and Titarchuck et al. (2001) to explain the presence of NBOs in Z sources.
Titarchuk et al. (2001) suggests a model where the normal branch oscillations are considered to be viscous 
oscillations of a spherical shell around the NS surface. Based on equation 1 from the
Titarchuk et al. (2001) as given below, we estimated the size (L$_{s}$) of this shell based on our obtained NBO frequencies.
\begin{equation}
L_{s} = \frac{f \nu_{s}}{\nu_{ssv}}
\end{equation}
$\nu$$_{ssv}$ is the spherical shell viscous frequency, f is 0.5 for the stiff and 1/2$\pi$ for the free boundary conditions in the transition layer, $\nu$$_{s}$ is the sonic velocity which 
can be estimated based on the equation given by Hasinger (1987) as, 
\begin{equation}
\nu_{s}= 4.2 \times 10^{7} R_{6}^{-1/4} (\frac{M}{M_{\odot}} \frac{L}{L_{Edd}})^{1/8} cm \; s^{-1}
\end{equation}
Here R$_6$ is the neutron star radius in units of 10$^6$ cm. 
For a 7 Hz NBO with luminosity obtained from the spectral model, the size L$_s$ was estimated to be $\sim$ 32 km for f=0.5, which is within the estimates of the size determined for the boundary layer. 
It is difficult to differentiate between these two structures (transition layer and boundary layer) as they have very similar spectral and temporal properties and the context of invoking these two structures is simply to constrain their sizes.
Previously, Homan et al. (2001) has also detected NB0s of 6.3--7 Hz frequency and found FBOs ranging from 13.9--23.1 Hz. Most recently, 
Agrawal et al. (2020), using AstroSat LAXPC data, observed a 7.4 Hz NBO although FBOs were not detected in the FB.
Now, considering the same mechanism causes FBOs, L$_{s}$ was found to be $\sim$ 11 km for a 20 Hz FBO.

\section{Conclusion}

 Using energy dependent CCF studies we have detected anticorrelated lags of the order few 100 s using 
   Astrosat LAXPC and SXT light curves. We conclude that the detected lags 
   are readjustment timescales of corona in the inner region of the accretion disk. Moreover only power law index varied from top to bottom of the 
   normal branch, suggesting that corona was changing. Based on CCF lags the Compton cloud / corona size is 
   found to be around 27--46 km. 
   The inner region between the accretion disk and the NS surface could host a Boundary Layer or a transition shell causing
    the NBO, or there could also be a puffing up of the inner disk due to radiation pressure. The exact structural setup in this
   region still remains inconclusive. We therefore estimate the sizes of each of these possible components,
   using various methods as discussed above and this indicate towards a component of size varying from 30--100 km. 
   
   Broadband spectral (0.8--50 keV) results put a strong constrain on the inner disk radius, 
   R$_{eff}$ = 12--16 km (5.7--8.0 R$_{g}$) for GX 17+2 in the normal branch and these values closely matches with other works 
   (Cackett et al 2010; Ludlam et al. 2017). This strongly indicates that the disk is close to the ISCO and the disk front is not 
   traversing along the normal branch supporting our assumption that corona is only changing during detected lags. 

Size estimation of corona/sub-keplerian flow from different methods as discussed above indicates that a vertical structure of 
few tens of km is needed to explain the detected lags, NBOs and spectra of GX 17+2. 
Similar studies especially covering the soft spectral energy domain along with hard spectral energy domain are 
essential to constrain the accretion disk geometry in Z sources.
Hence future observations from Astrosat (SXT and LAXPC) and Nustar are required in order to understand the nature of lags in Z sources 
and their connection to the corona or boundary layer.

\section*{Acknowledgements}
K.S. acknowledges  the  financial  support of ISRO under AstroSat  archival  Data  utilization  program.  This   
publication  uses data  from  the  AstroSat  mission  of  the Indian  Space  Research Organisation (ISRO), archived at the Indian 
Space Science DataCentre (ISSDC). M.S. acknowledges the financial support from DST-INSPIRE fellowship.
Authors sincerely acknowledge the contribution of the LAXPC and SXT instrument teams toward the development of the LAXPC and SXT
instruments onboard the AstroSat. This research has made use of the data collected from the AO cycle 3 of AstroSat observations.
This research work has used the data from the Soft X-ray Telescope (SXT) developed at TIFR,
Mumbai, and the SXT POC at TIFR is acknowledged for verifying and releasing the data via the ISSDC data archive
and also for providing the necessary software tools required for the analysis.
This work also uses data from the LAXPC instruments developed at TIFR, Mumbai and the LAXPC POC at
TIFR is thanked for verifying and releasing the data via the ISSDC data archive. Authors thank
the AstroSat Science Support Cell hosted by IUCAA and TIFR for providing the LAXPC
software which was used for LAXPC data analysis. We thank the Referee for providing valuable 
suggestions that have improved the quality of the paper.

\section*{Data Availability}
Data used in this work can be accessed through the Indian Space Science Data Center (ISSDC) website 
(https://astrobrowse.issdc.gov.in/astro\_archive/archive/Home.jsp) and is also available with the authors.

%%%%%%%%%%%%%%%%%%%%%%%%%%%%%%%%%%%%%%%%%%%%%%%%%%

\begin{figure*}
\centering
\includegraphics[width=0.5\textwidth, angle=0]{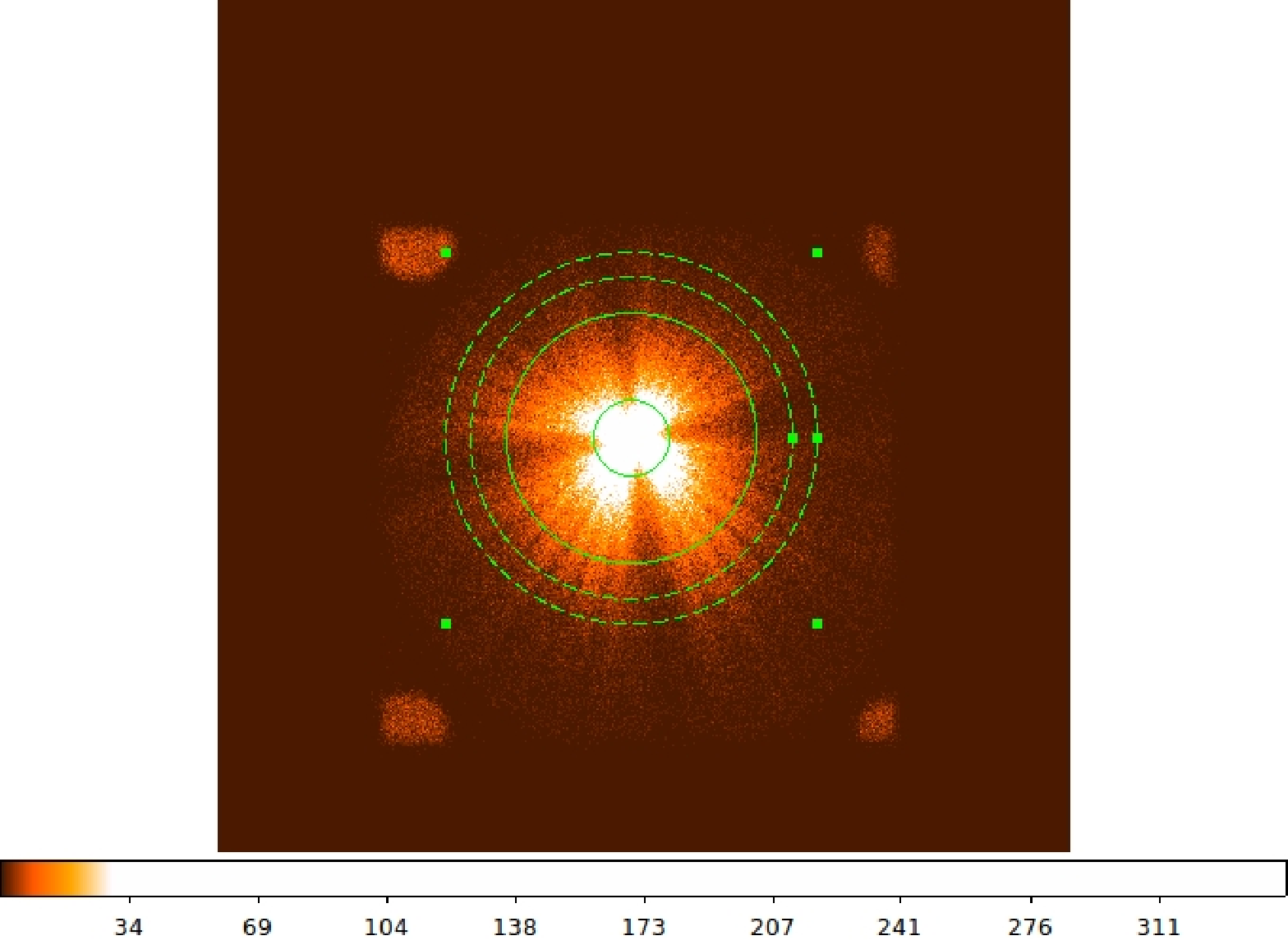}\\

\caption{ Top: SXT image of the source and the extraction regions used for source (solid annulus) and background (dashed annulus) extraction. Calibration
sources can be seen in the four corners of the image.}
\end{figure*}

\begin{figure*}
\centering
\includegraphics[height=0.5\textwidth, width=5cm, angle=270]{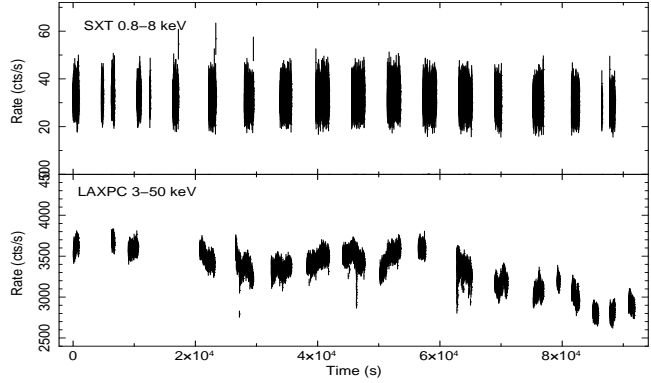}\\
\caption{ Top: SXT light curve in the 0.8-8 keV energy range. Bottom: Combined LAXPC 10 and 20 light curve for the entire 3-50 keV energy range. Both plotted
using a time bin of SXT resolution, 2.3775s.}
\end{figure*}

\begin{figure*}
\centering
\includegraphics[height=0.5\textwidth, width=5cm, angle=270]{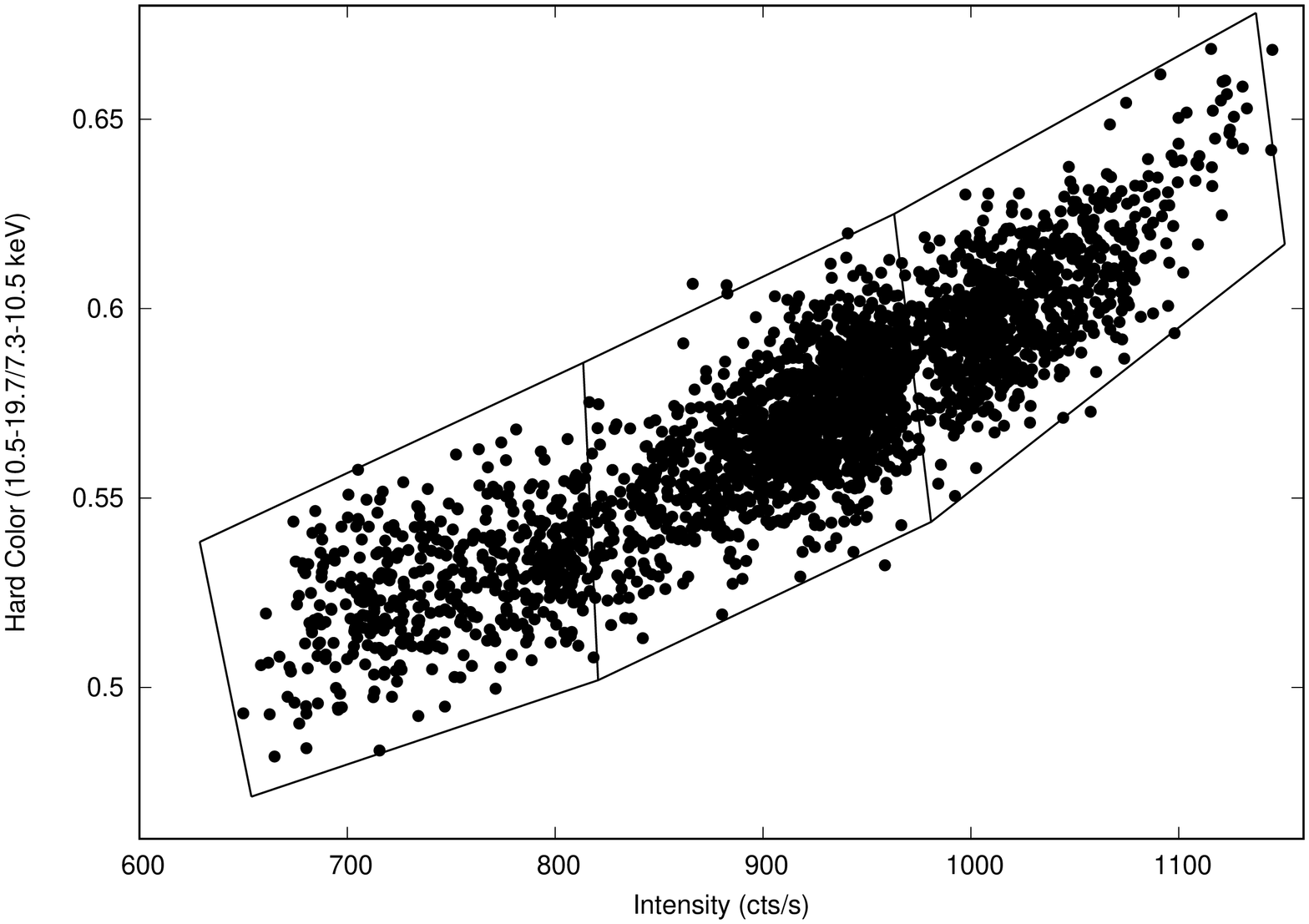}\\

\caption{ HID for GX 17+2 using AstroSat LAXPC observations. Hard colour is defined as  10.5-19.7/7.3-10.5 keV and Intensity is that in the 7.3-19.7 keV range. Boxes indicate the separations of HID into
upper (A), middle (B) and lower (C) sections.}
\end{figure*}

\begin{subfigures}
\begin{figure*}
%\begin{center}
\caption{ The background subtracted 30s bin LAXPC soft (3-5 keV) and hard X-ray (16--20 keV, 20--40 keV, 20--50 keV \& 20--60 keV) light curves (left panels) for which CCF lags are observed (right panels).
Energy bands used are mentioned in the light curves (left panel). Right panels show the cross correlation function (CCF) of each section of the light curve and shaded regions show the standard deviation 
of the CCFs. } 
\includegraphics[height=\textwidth, width=8cm, angle=270]{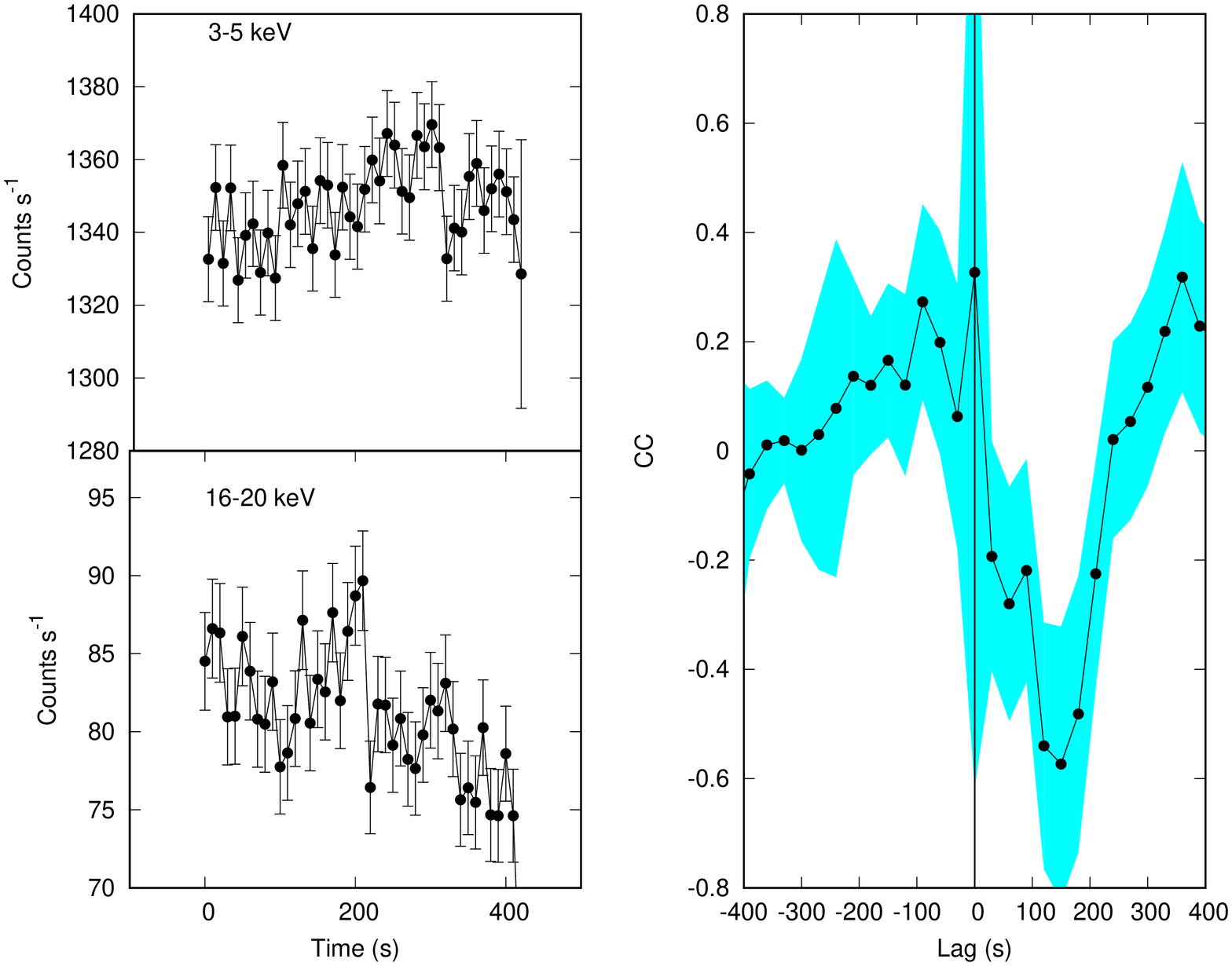} \\
\end{figure*}
\begin{figure*}
\caption{}
\includegraphics[height=\textwidth, width=8cm, angle=270]{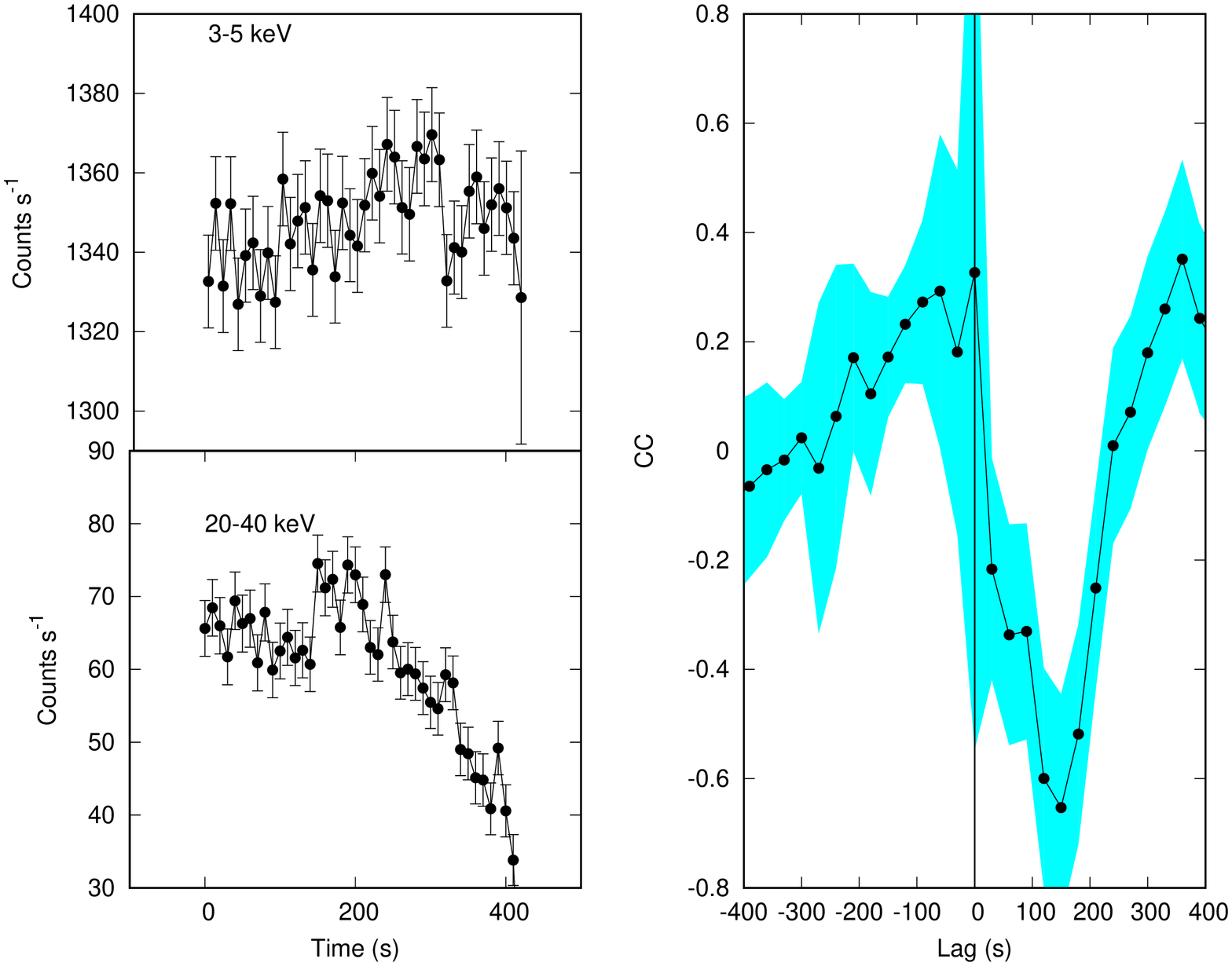}\\
\end{figure*}
\begin{figure*}
\caption{}
\includegraphics[height=\textwidth,width=8cm, angle=270]{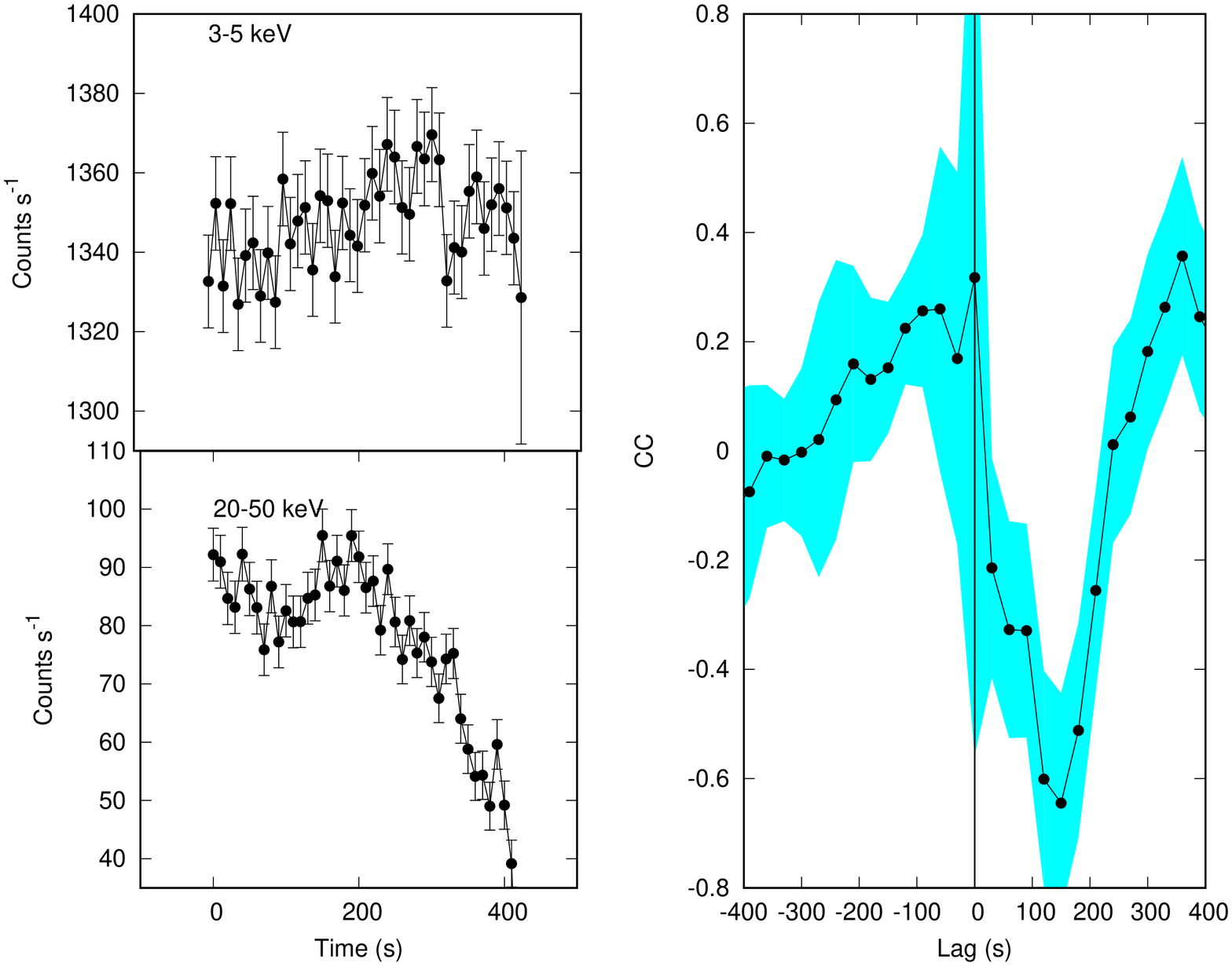}\\
\end{figure*}
\begin{figure*}
\caption{}
\includegraphics[height=\textwidth,width=8cm, angle=270]{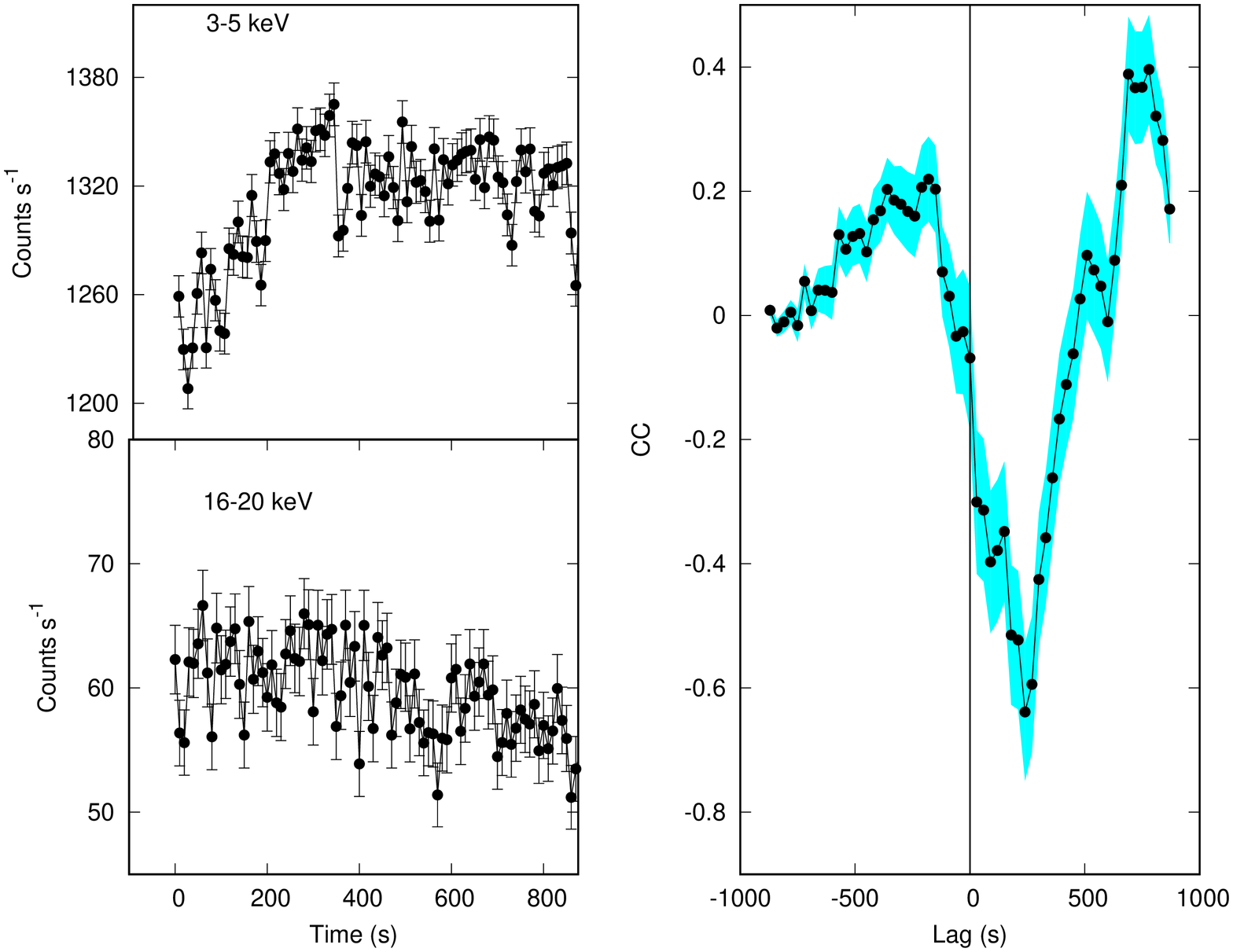}\\
\end{figure*}
\begin{figure*}
\caption{}
\includegraphics[height=\textwidth,width=8cm, angle=270]{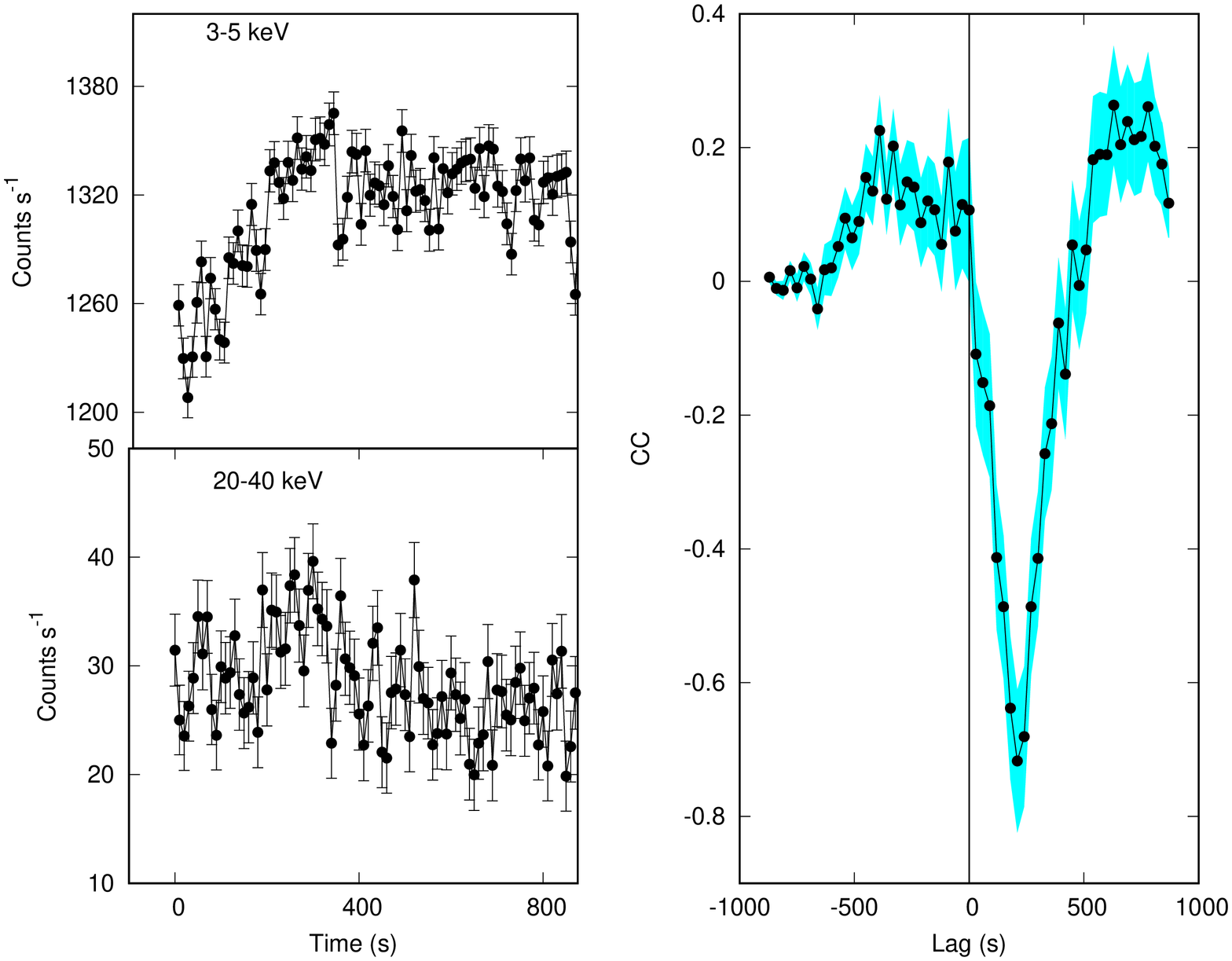}\\
\end{figure*}
\begin{figure*}
\caption{}
\includegraphics[height=\textwidth, width=8cm, angle=270]{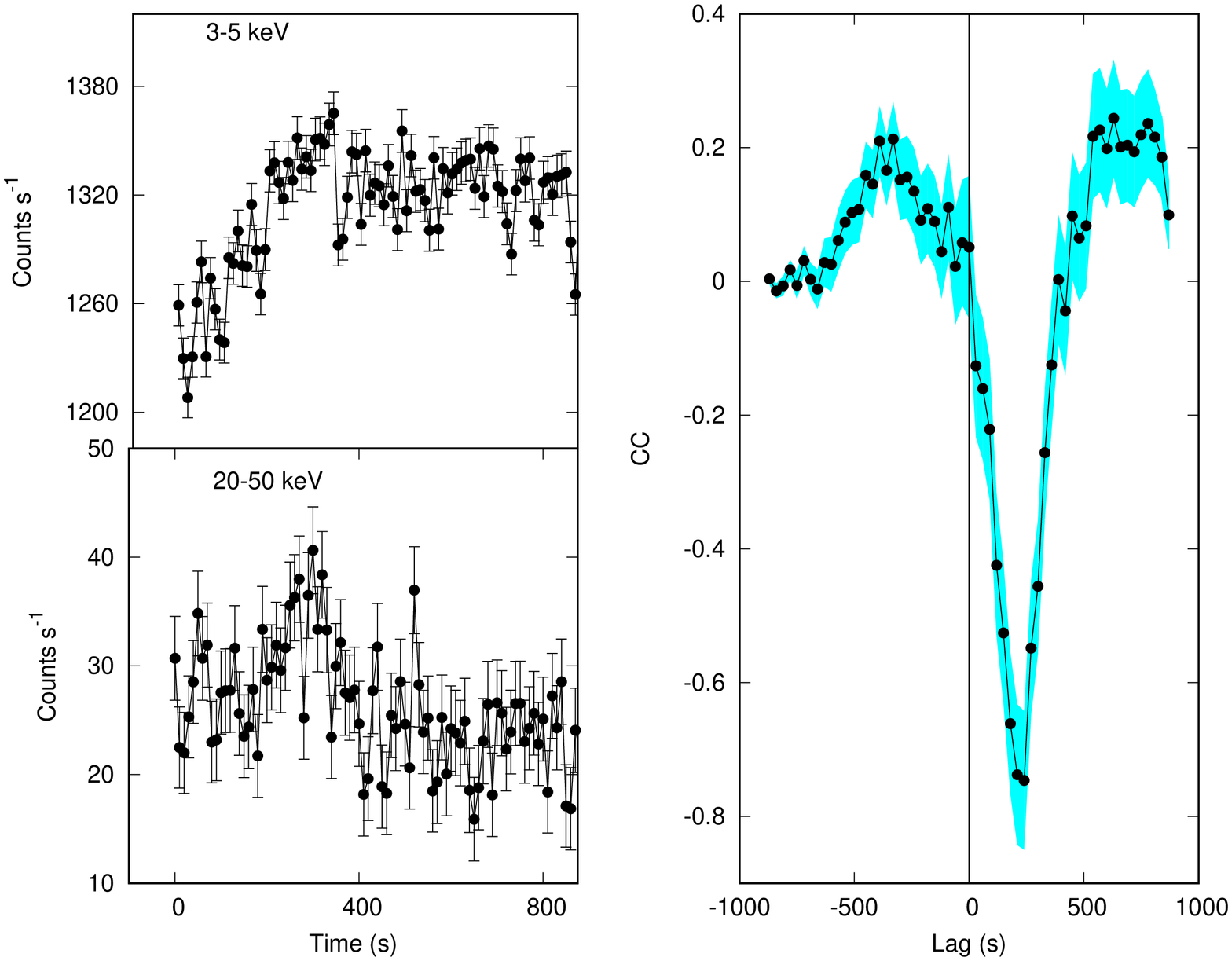}\\
\end{figure*}
\end{subfigures}

\begin{subfigures}
\begin{figure*}
%\begin{center}
\caption{The background subtracted SXT soft (0.8-2 keV) and LAXPC hard X-ray (10--20 keV, 16--20 keV, 20--40 keV, 20--50 keV \& 20--60 keV) light curve (left panels) for which CCF lag is observed (right panels).
Energy bands used are mentioned in the light curves (left panel). Right panels show the cross correlation function (CCF) of each section of the light curve and shaded regions show the standard deviation 
of the CCFs. }
\includegraphics[height=\textwidth, width=8cm, angle=270]{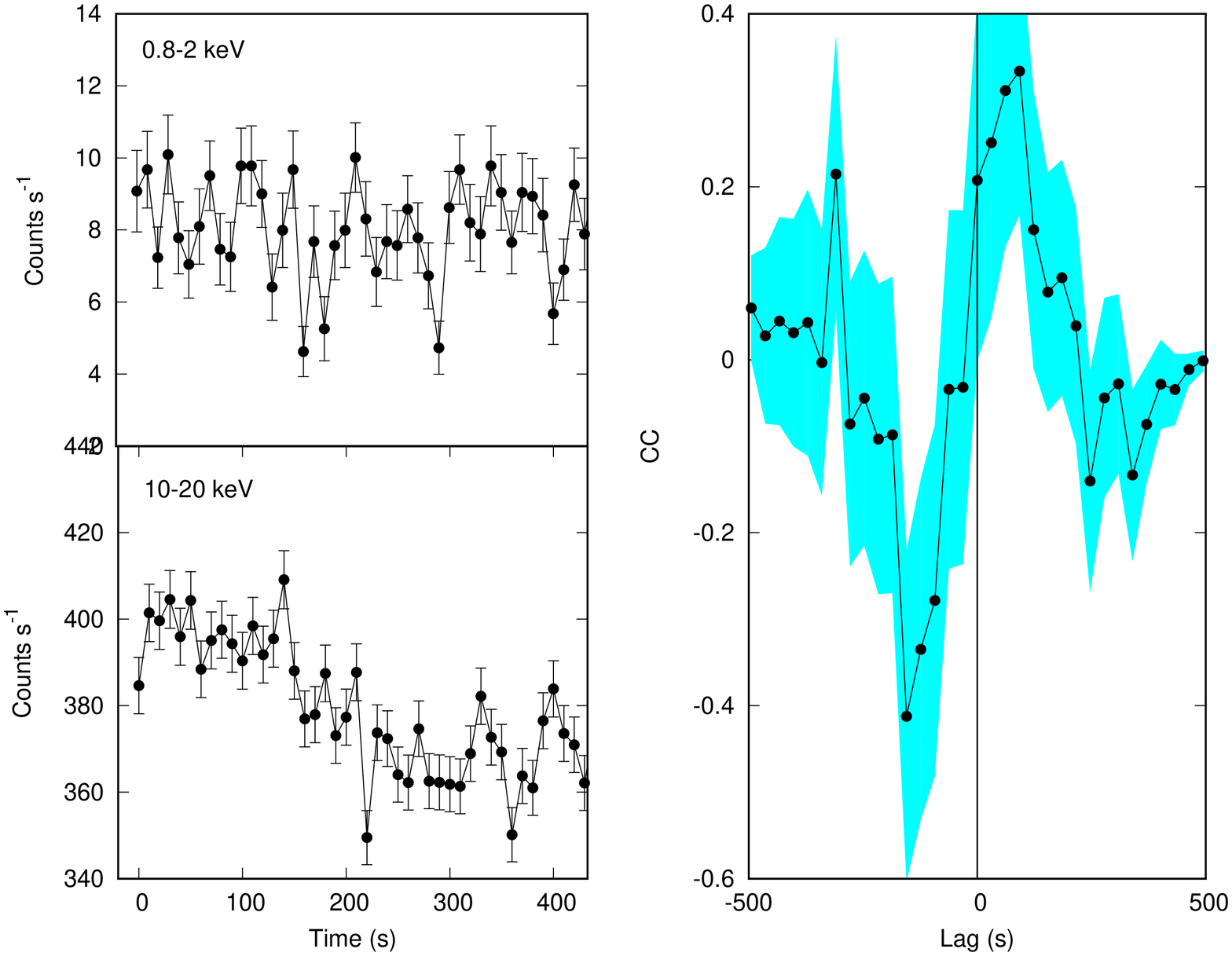} \\
\end{figure*}
\begin{figure*}
\caption{}
\includegraphics[height=\textwidth, width=8cm, angle=270]{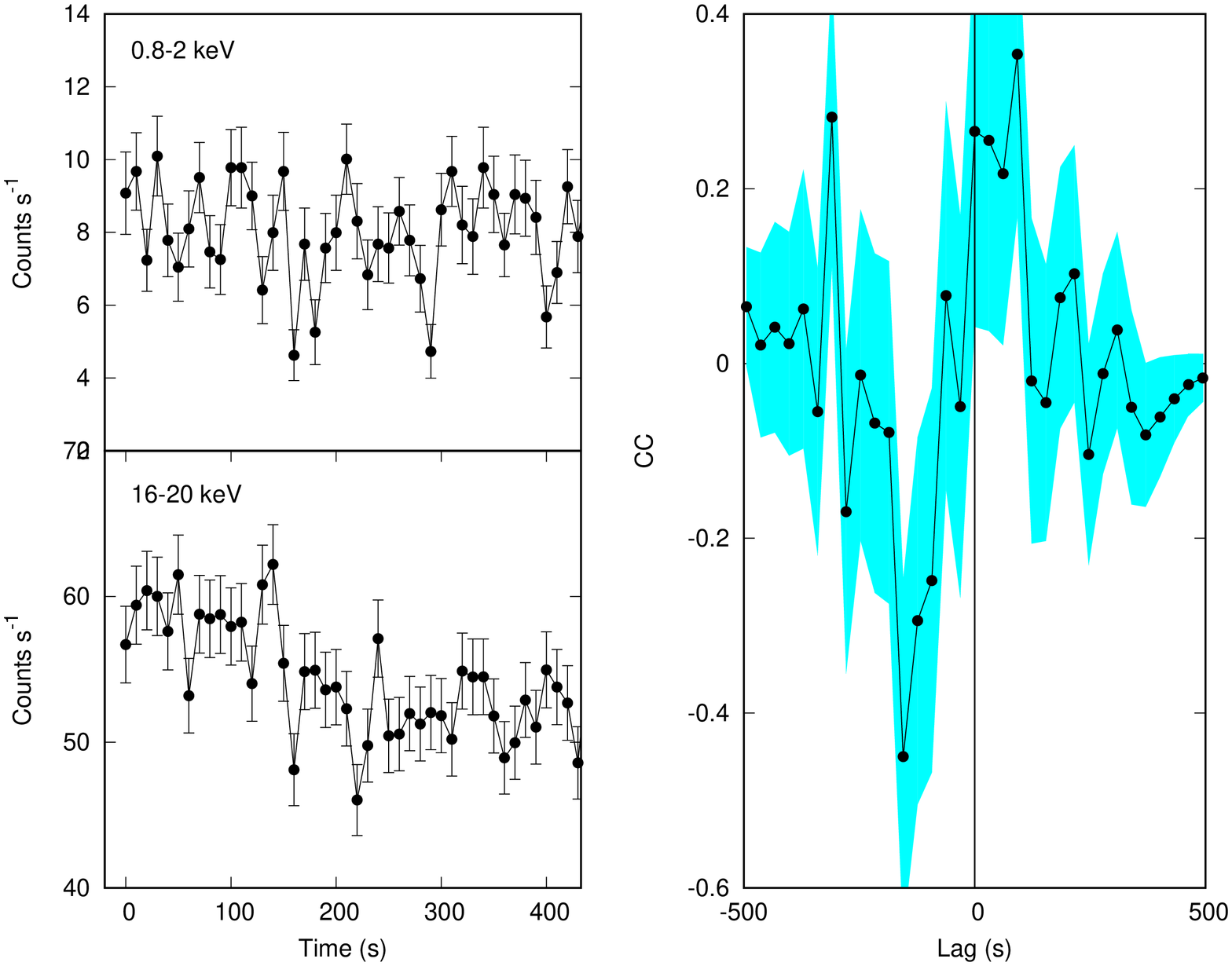} \\
\end{figure*}
\begin{figure*}
\caption{}
\includegraphics[height=\textwidth, width=8cm, angle=270]{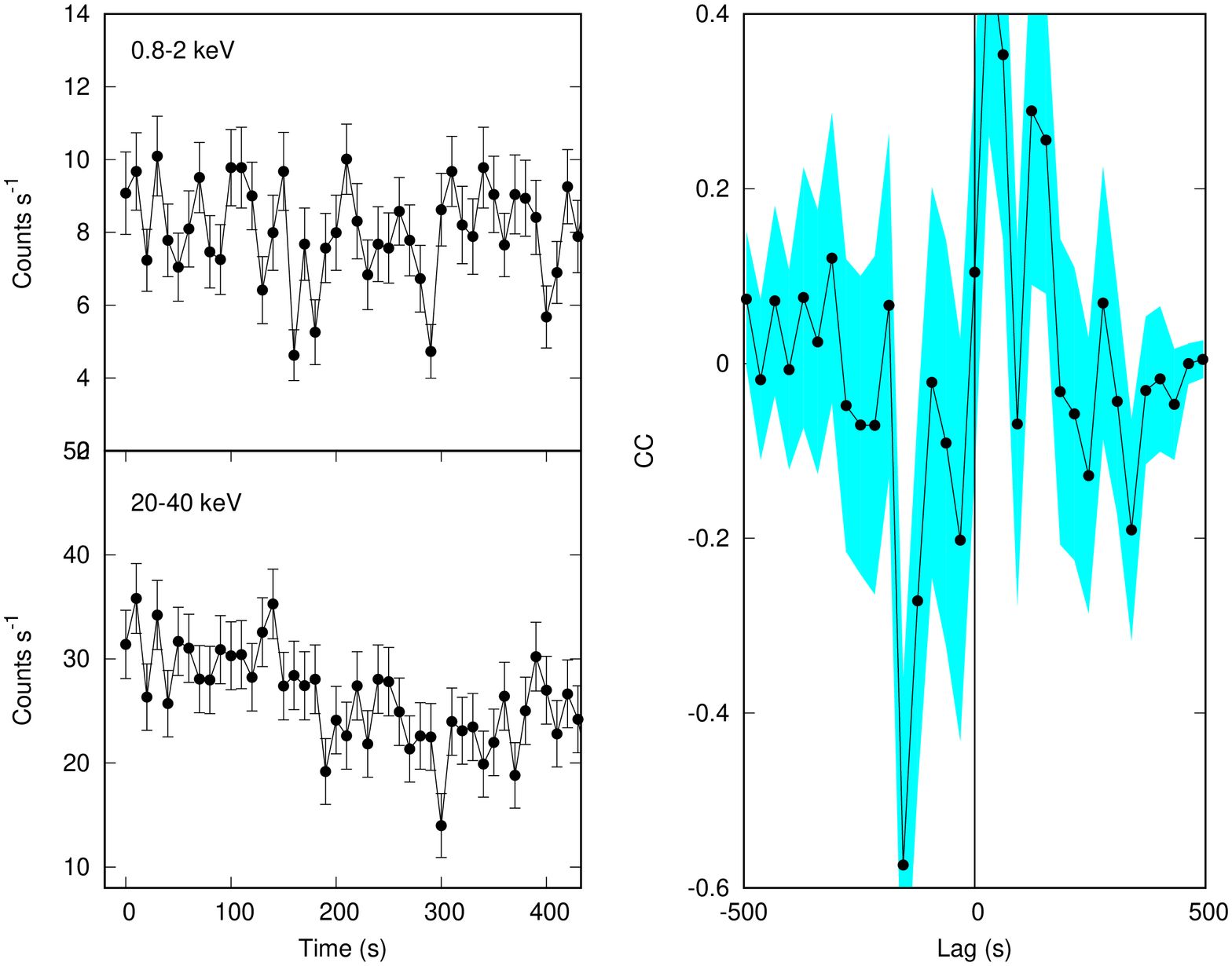} \\
\end{figure*}
\begin{figure*}
\caption{}
\includegraphics[height=\textwidth, width=8cm, angle=270]{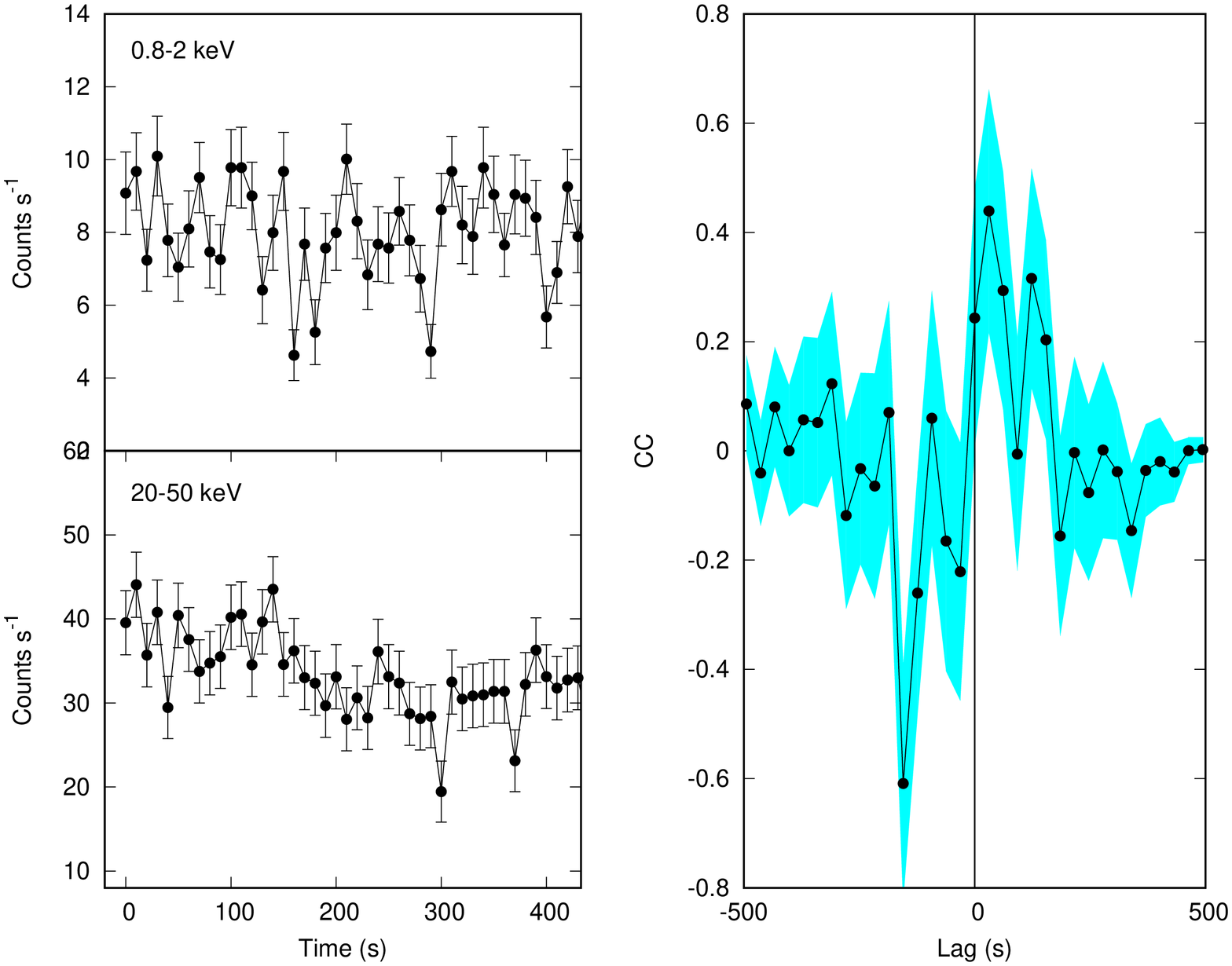} \\
\end{figure*}
\end{subfigures}

\begin{subfigures}
\begin{figure*}
\caption{ QPOs seen in the last three segments of the light curve in 3-20 keV energy range for GX 17+2 using AstroSat LAXPC observations along with their residuals in the bottom panel
after fitting a lorentzian and powerlaw model to the PDS.}
\includegraphics[width=0.4\textwidth, angle=270]{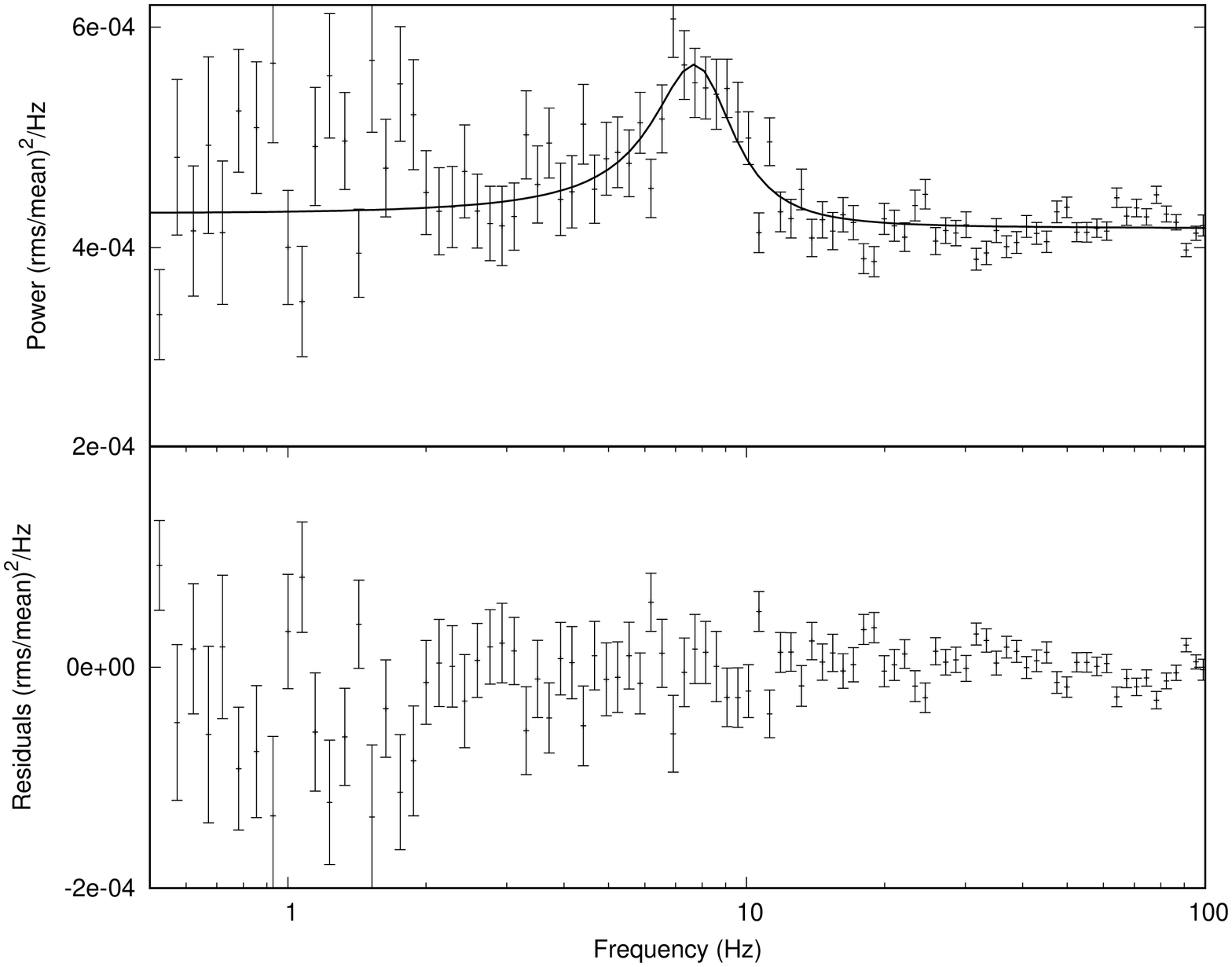}\\
\end{figure*}
\begin{figure*}
\caption{}
\includegraphics[width=0.4\textwidth, angle=270]{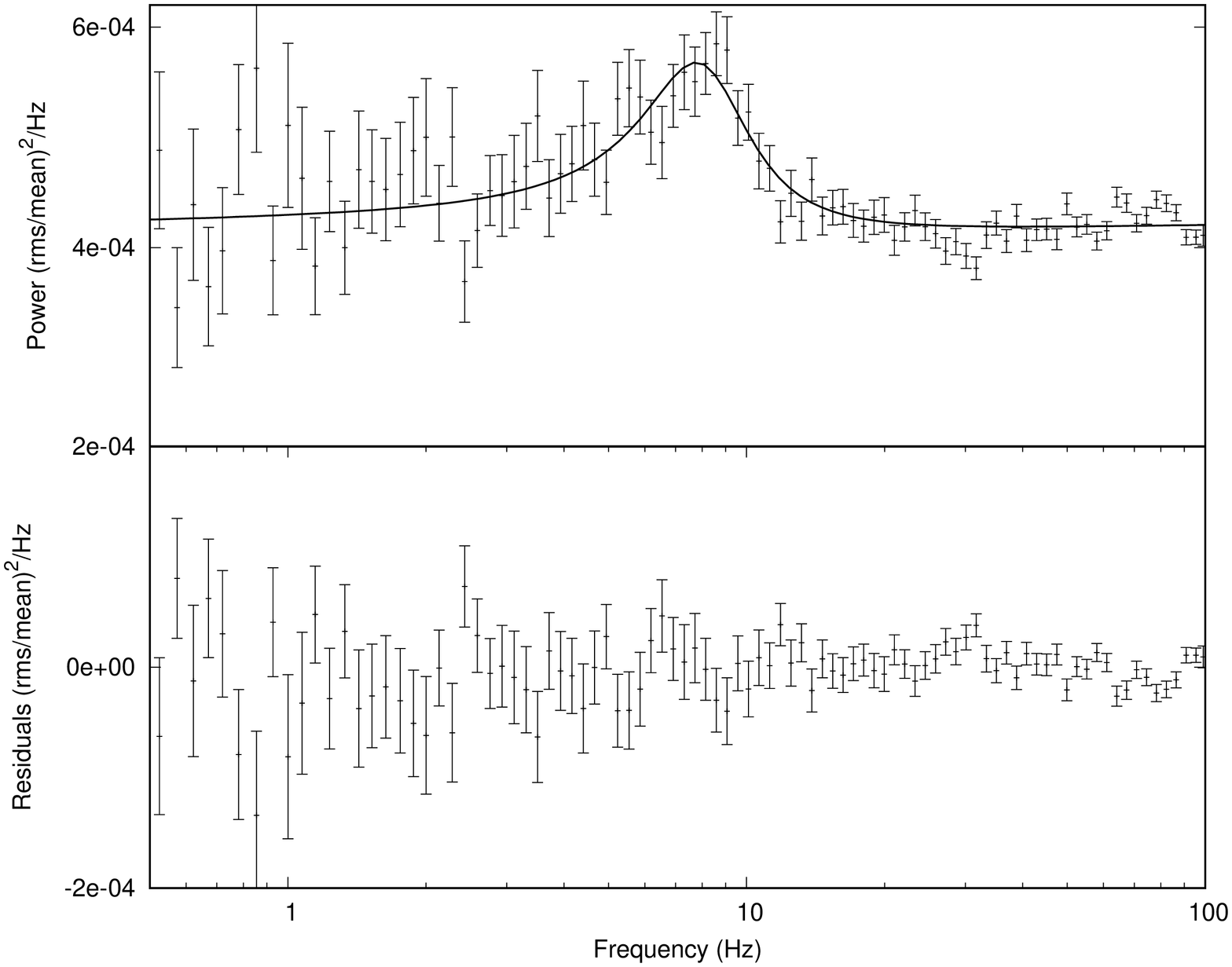}\\
\end{figure*}
\begin{figure*}
\caption{}
\includegraphics[width=0.4\textwidth, angle=270]{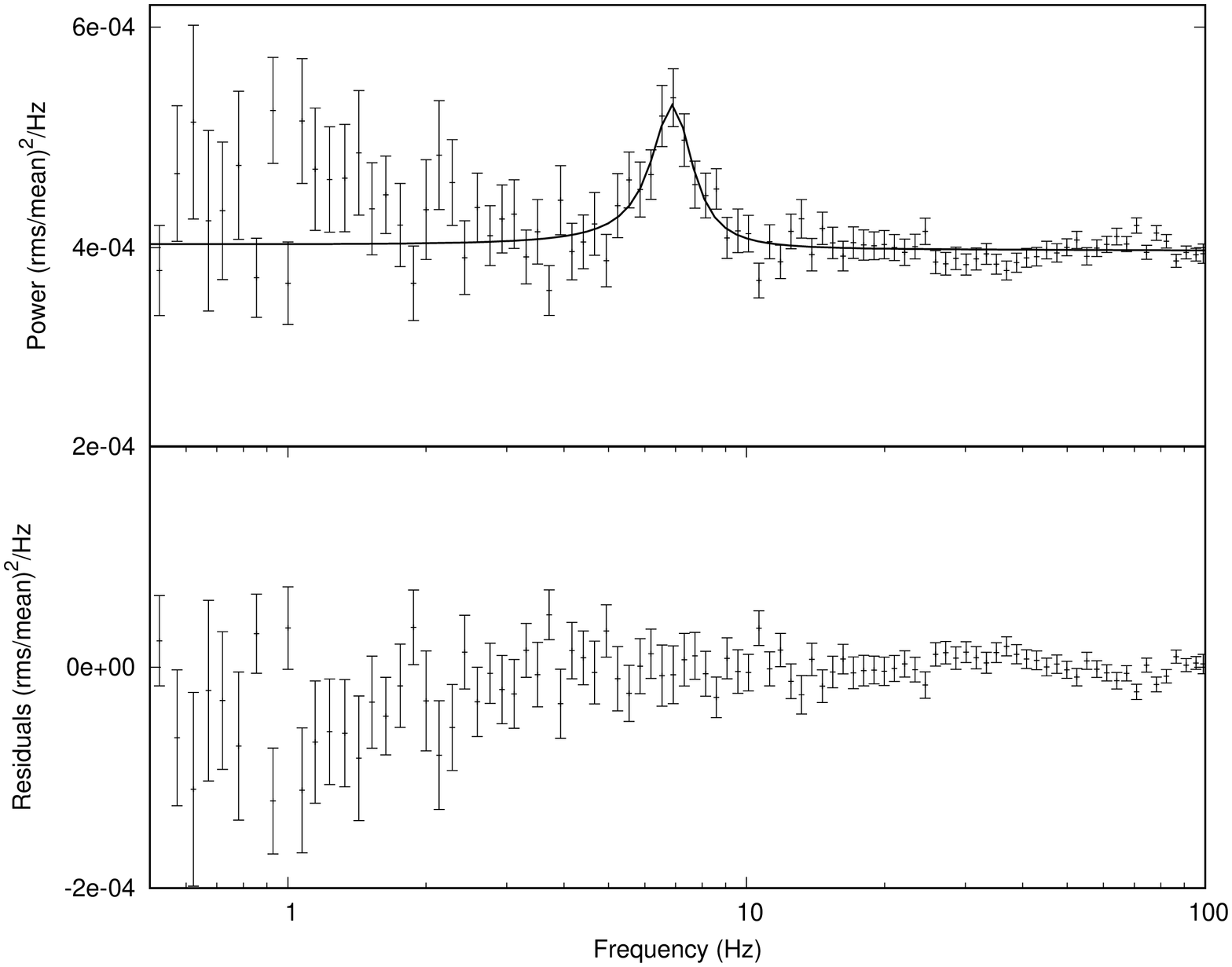}\\
\end{figure*}
\end{subfigures}

\begin{figure*}
\caption{Dynamic Power Density spectrum for the entire lightcurve in 3-20 keV energy range.}
\includegraphics[height=10cm,width=10cm, angle=270]{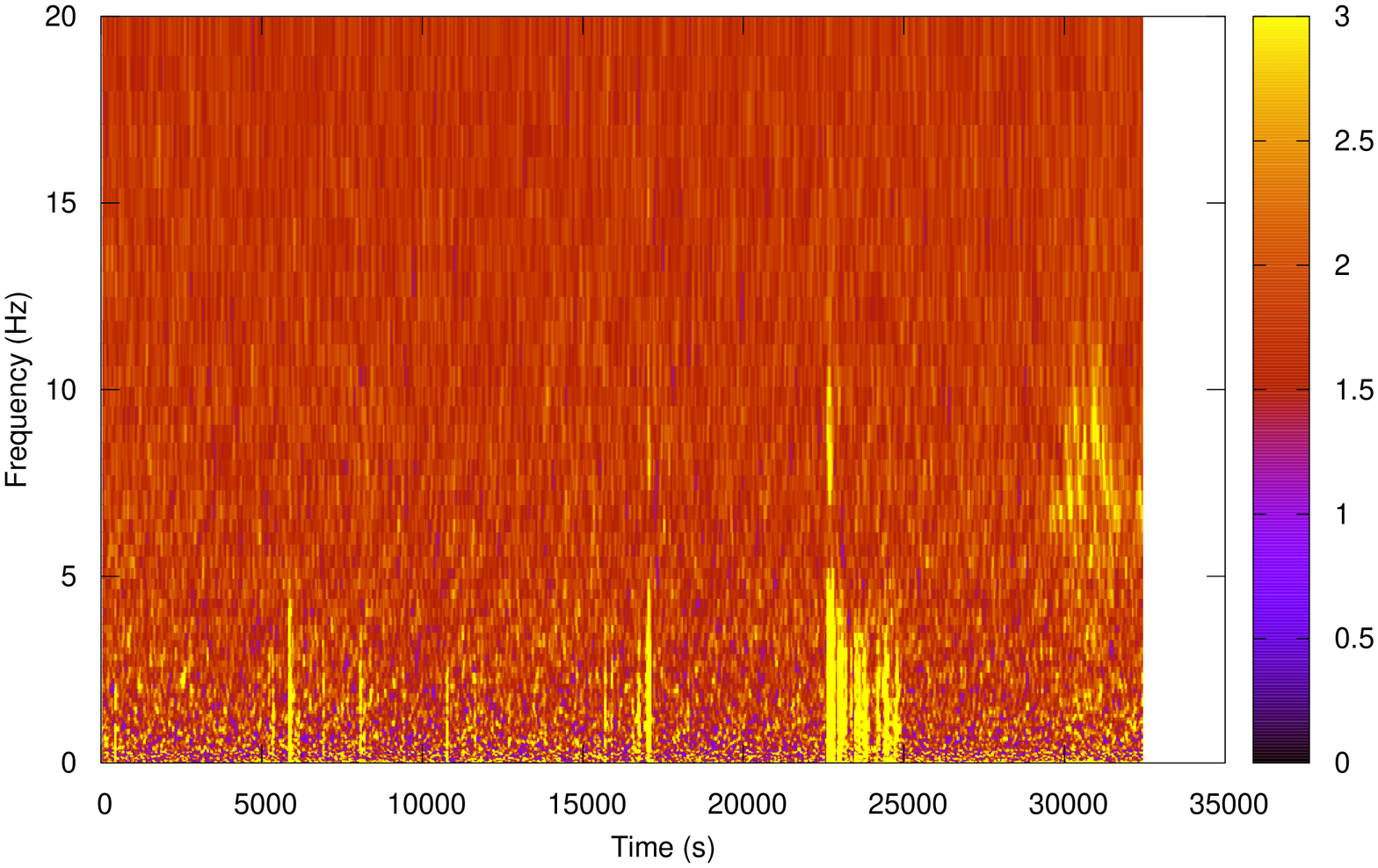}\\

\end{figure*}

\begin{subfigures}
\begin{figure*}
%\begin{center}
\caption{$\Delta$$\chi$$^2$ (=$\chi$$^2$-$\chi$$^2$$_{min}$) vs inclination angle for the best fit model (Table 3) (section 1).}
\includegraphics[height=0.4\textwidth,angle=270]{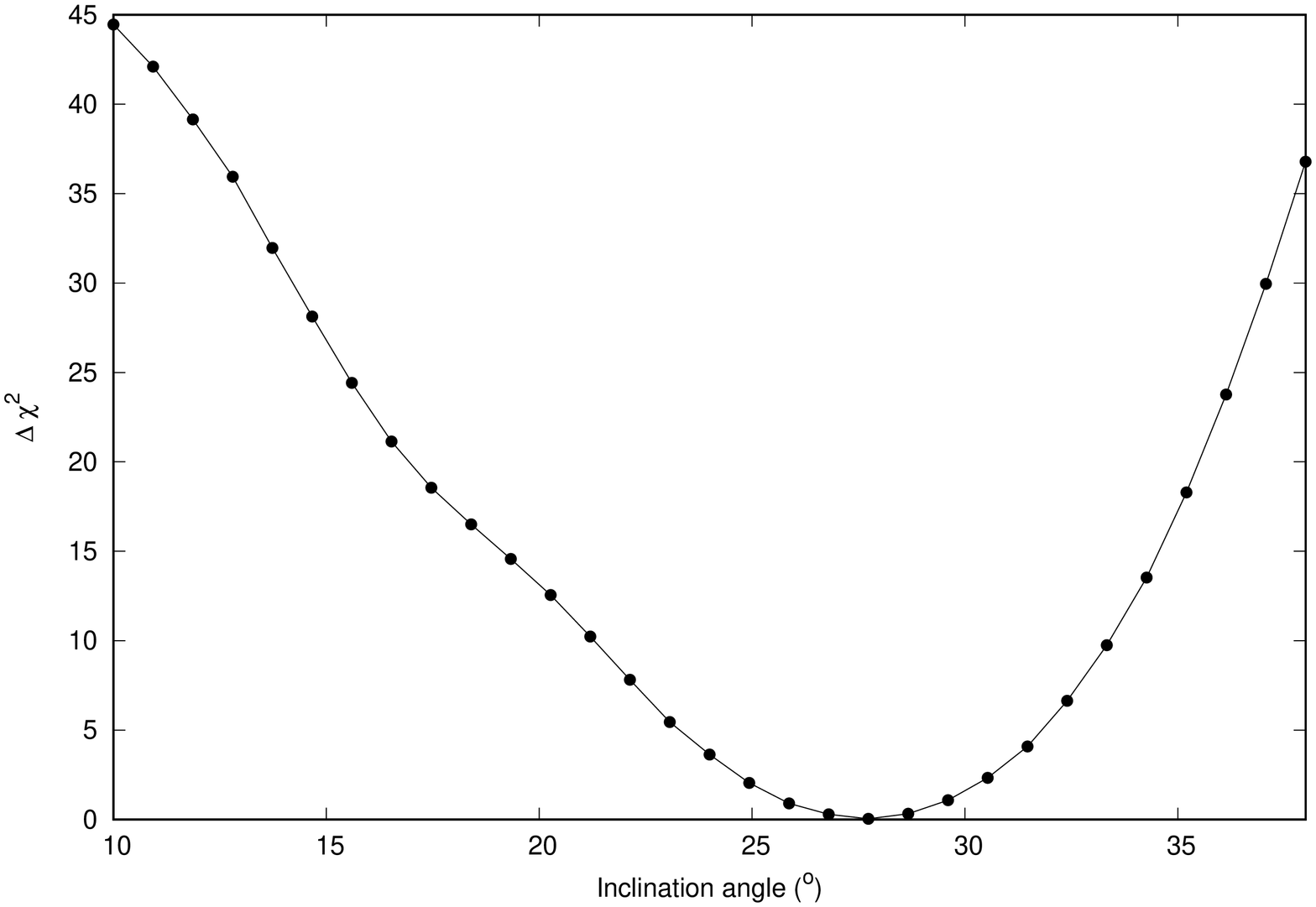} 
\end{figure*}
\begin{figure*}
\caption{$\Delta$$\chi$$^2$ (=$\chi$$^2$-$\chi$$^2$$_{min}$) vs  R$_{in}$ for the best fit model (Table 3) (section 1).}
\includegraphics[height=0.4\textwidth, angle=270]{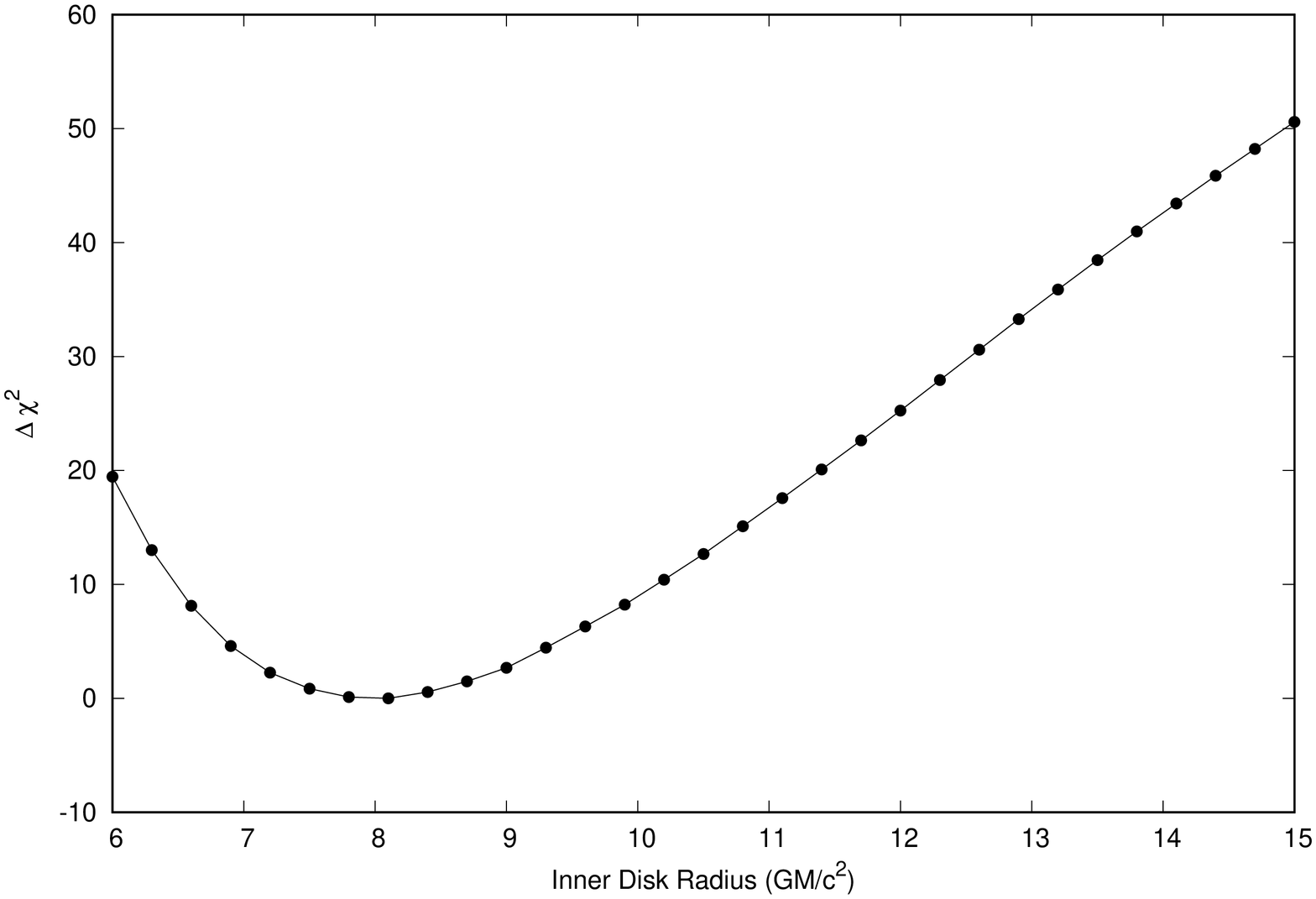} 
\end{figure*}
\end{subfigures}

\begin{subfigures}
\centering
\begin{figure*}
\caption{The SXT+LAXPC joint spectral fit for the the {\it Diskbb + rdblur*bbrefl + Gaussian (Xe) + Power-law} model for sections A. The top panel gives
the unfolded spectra (thick line) with the component models (dashed lines) and the bottom panel gives the residuals obtained from the fit. }
\includegraphics[height=0.5\textwidth,angle=270]{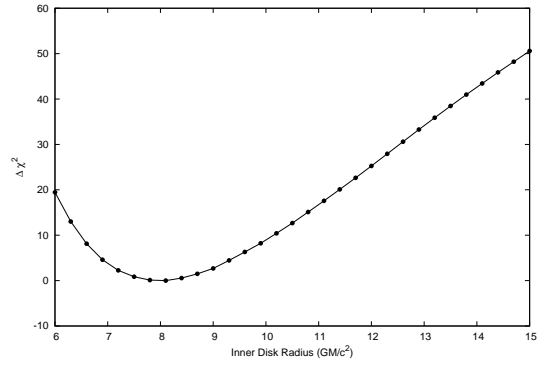} \\
\end{figure*}
\begin{figure*}
\caption{Similar figure for section B}
\includegraphics[height=0.5\textwidth, angle=270]{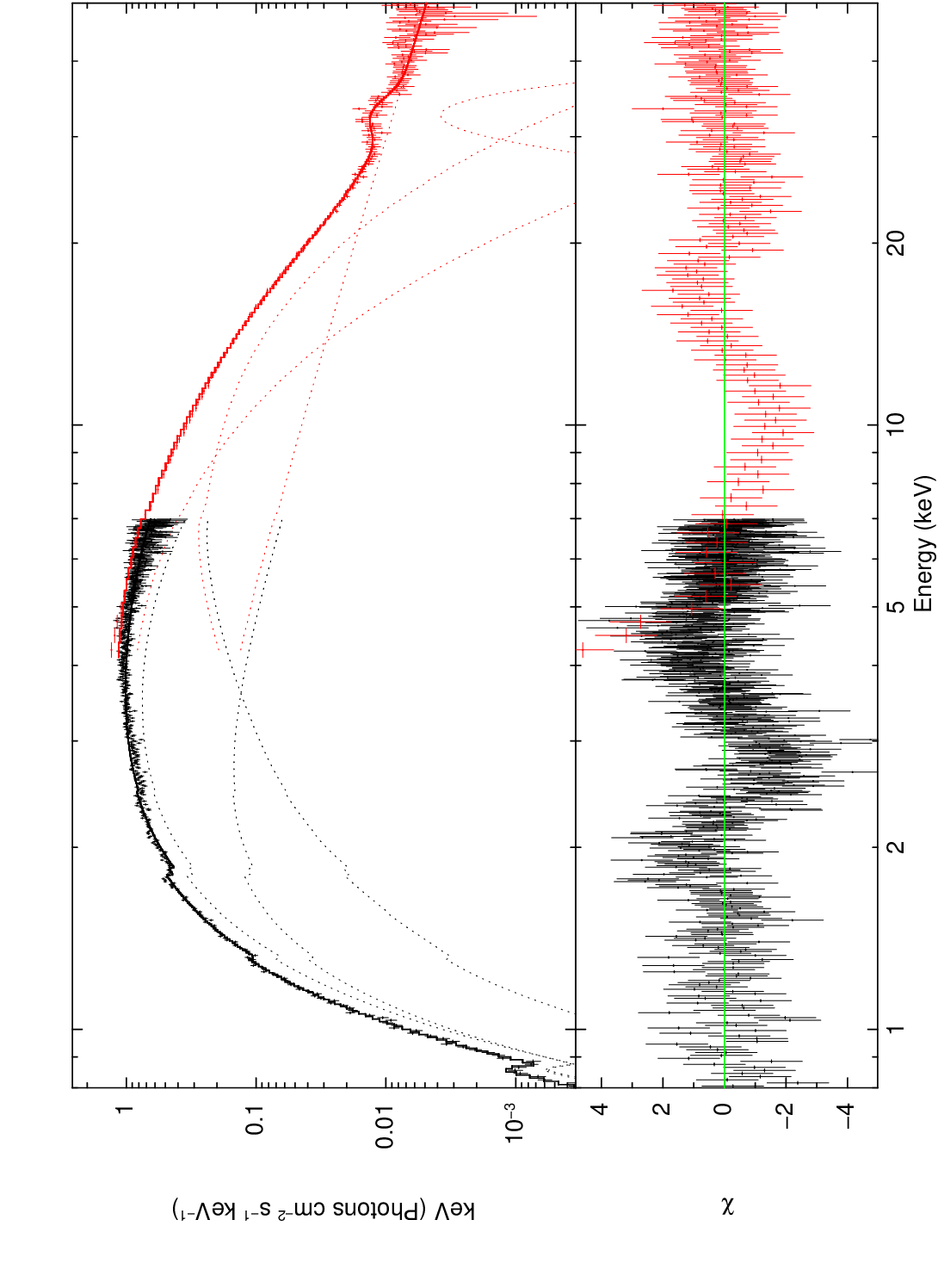} \\
\end{figure*}
\begin{figure*}
\caption{Similar figure for section C}
\includegraphics[height=0.5\textwidth, angle=270]{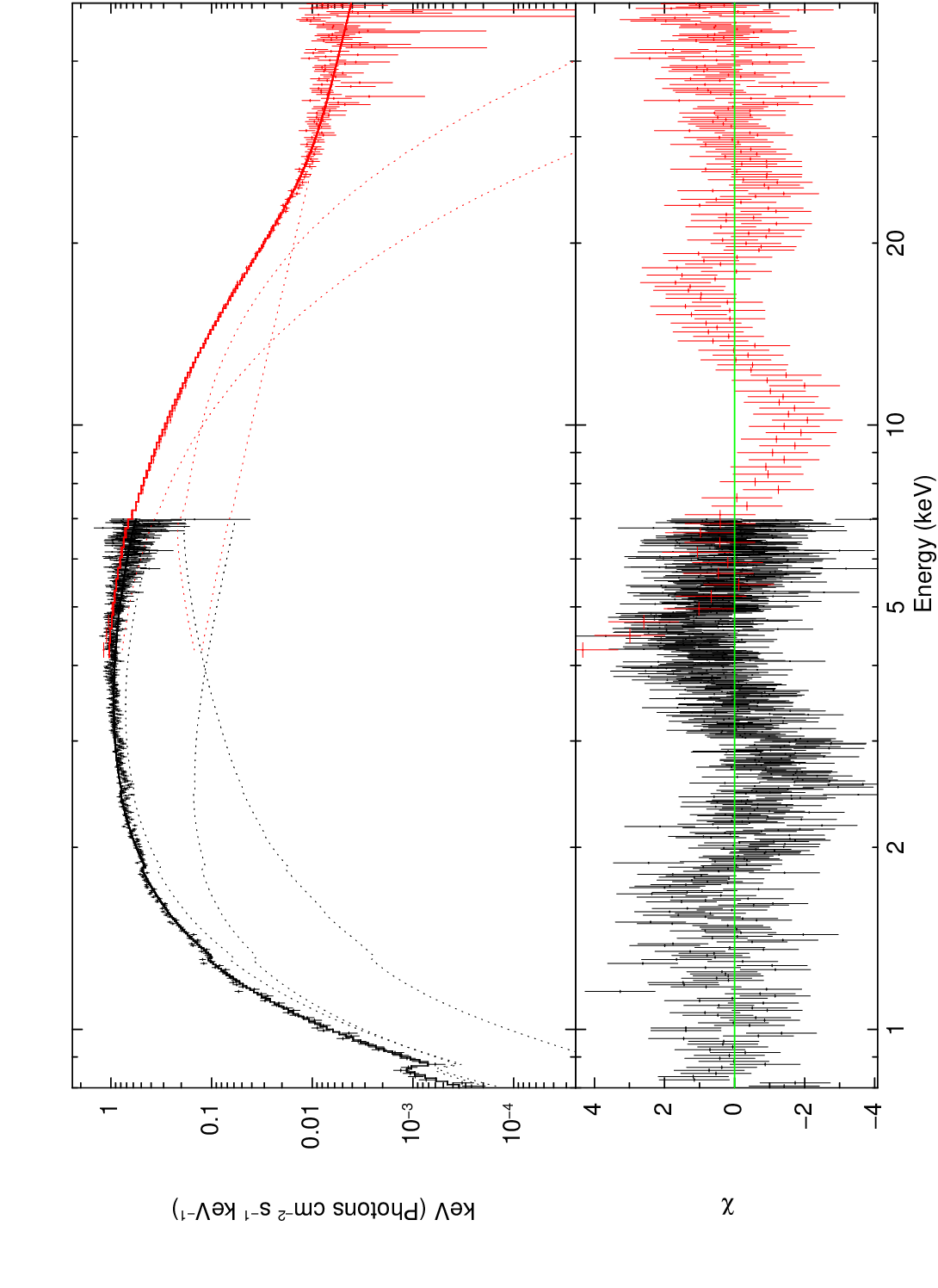} \\
\end{figure*}
\end{subfigures}
\clearpage

\begin{table}
\begin{minipage}[t]{\columnwidth}
\scriptsize
\caption{CCF of LAXPC and SXT lightcurves. Here I, II and III identifies the different light curve sections which exibited lags.}
\subcaption{ CCF of LAXPC soft vs. hard light curves. }
\centering
\begin{tabular}{|c|cc|cc|cc|cc}
\toprule
&\multicolumn{2}{|c|}{3-5 vs. 16-20 keV} & \multicolumn{2}{c|}{3-5 vs. 20-40 keV}&\multicolumn{2}{c|}{3-5 vs. 20-50 keV}\\ 
\midrule
& Lag (s)  & CC  & Lag (s)  &CC & Lag (s) & CC \\ 
\hline
\multicolumn{1}{|c|}{I} & 217 $\pm$ 20 & -0.64 $\pm$ 0.11 & 215 $\pm$ 13 & -0.72 $\pm$ 0.11 & 212 $\pm$ 10 & -0.75 $\pm$ 0.10 \\
\multicolumn{1}{|c|}{II} & 139  $\pm$ 31 & -0.57 $\pm$ 0.25 & 135 $\pm$ 24 & -0.65 $\pm$ 0.21 & 135$\pm$ 23 & -0.64 $\pm$ 0.20  \\
\bottomrule 
\end{tabular}%
\\
\bigskip 
\subcaption{ CCF of SXT soft vs. LAXPC hard light curves.}
\begin{tabular}{|c|cc|cc|cc|cc|cc}
\toprule
&\multicolumn{2}{|c|}{0.8-2 vs. 10-20 keV}&\multicolumn{2}{c|}{0.8-2 vs. 16-20 keV} & \multicolumn{2}{c|}{0.8-2 vs. 20-40 keV}&\multicolumn{2}{c|}{0.8-2 vs. 20-50 keV} \\ 
\midrule
%\cmidrule(lr){2-4}\cmidrule(lr){5-7}\cmidrule(lr){8-9}\cmidrule(lr){10-11}
& Lag (s)  & CC  & Lag (s)  &CC & Lag (s) & CC & Lag (s)  &CC   \\ 
\hline
\multicolumn{1}{|c|}{III} & -133 $\pm$ 34 & -0.41 $\pm$ 0.19 & -139 $\pm$ 35 & -0.45 $\pm$ 0.21 & -141 $\pm$ 15 & -0.57 $\pm$ 0.21 &-141 $\pm$ 15 & -0.61 $\pm$ 0.22  \\
\bottomrule
\end{tabular}%

\end{minipage}
\end{table}%
\clearpage

\begin{table}{}
\begin{minipage}[t]{\columnwidth}
\scriptsize
\caption{Best-fit spectral parameters for the upper, middle and lower sections of the HID which the letters A, B and C respectively represent using the
 {\it Diskbb + Bbody + Gaussian(Fe) +  Gaussian (Xe) + Powerlaw} model. Energy of Xe line here is $\sim$ 32.4 keV.
 The subscript BB represents the bbody model and dBB represents Diskbb model.
The flux in units of 10$^{-8}$ ergs cm$^{-2}$ s$^{-1}$ is calculated in the energy band 0.8--50 keV. Errors are quoted at a 90\% confidence level.
Luminosity is in units of 10$^{38}$ erg s$^{-1}.$ assuming the distance 13 kpc for GX 17+2.}
\label{tab2}
\centering
\begin{tabular}{ccccccccc}
\hline

\hline
%\hline
\hline
%Parameters&\multicolumn{2}{c}{A}&\multicolumn{2}{c}{B}&\multicolumn{2}{c}{C}&\multicolumn{2}{c}{D}\\
Parameters&A&B&C&\\
\hline
$kT_{in}$ (keV)\footnote{Temperature of the Diskbb model.} &1.97$_{-0.09}^{+0.09}$ &1.98$_{-0.06}^{+0.07}$ &1.93$_{-0.06}^{+0.06}$ \\ \\
$N_{dBB}$\footnote{Normalization of the Diskbb model.}& 37.72$_{-5.38}^{+6.06}$&38.60$_{-4.06}^{+4.51}$&39.56$_{-4.15}^{+4.72}$\\ \\
$R_{eff}$($25^{\circ})$\footnote{Effective radius obtained by using Diskbb Normalizaton and the spectral corrections 1.18--1.64.}& 9.8 - 13.8 km & 10.0 - 13.9 km& 10.1 - 14.1 km \\ \\
$kT_{BB}$ (keV)\footnote{Temperature of the BB model.} &2.51$_{-0.08}^{+0.10}$ &2.61$_{-0.09}^{+0.11}$ &2.64$_{-0.13}^{+0.16}$ \\ \\
$N_{BB}$\footnote{Normalization of the BB model.}& 0.033$_{-0.006}^{+0.006}$&0.024$_{-0.004}^{+0.005}$&0.016$_{-0.003}^{+0.003}$\\ \\
$E_{Fe}$ (keV) \footnote{Line Energy of the Gaussian model for Iron line.}&6.7 &6.7 &6.7 \\ \\
$\sigma_{Fe}$\footnote{Line width of the Gaussian model for Iron line.}&0.5 &0.5&0.5\\ \\
$EqWidth_{Fe}$\footnote{Equivalent width of the Gaussian model for Iron line in units of eV.}&71 $_{-50}^{+46}$ & 83 $_{-51}^{+50}$ & 117 $_{-54}^{+47}$\\ \\
$N_{Fe}$\footnote{Normalization of the Gaussian model for Iron line.}&0.007&0.007&0.007 \\ \\
$\Gamma_{pl}$\footnote{Power-law index.}&2.31$_{-0.03}^{+0.03}$&2.41$_{-0.02}^{+0.02}$&2.46$_{-0.03}^{+0.03}$ \\ \\
$N_{pl}$\footnote{Normalization of the PL model.}&0.94$_{-0.09}^{+0.09}$ &1.00$_{-0.07}^{+0.07}$&0.99$_{-0.09}^{+0.09}$\\ \\
Total flux&2.41&2.29&2.05\\ \\
BB flux &0.39&0.28&0.18\\ \\
Powerlaw flux& 0.46&0.41&0.38\\ \\
$R_{sp}$\footnote{Spherization radius}&42.44 km&40.33 km &36.10 km \\ \\
$R_{BL}$\footnote{Boundary Layer radius}&70.62 km&65.98 km&57.24 km \\ \\
L$_{0.8-50 keV}$&4.85&4.61&4.13 \\ \\
$\chi^{2}$/dof&995/788&1017/788&1071/788\\ \\

\hline
\end{tabular}
\end{minipage}
\end{table}

\begin{table}
\begin{minipage}[t]{\columnwidth}
\scriptsize
\caption{Best-fit spectral parameters for the upper, middle and lower sections of the HID which the letters A, B and C respectively represent using the 
{\it Diskbb + rdblur*bbrefl + Gaussian (Xe) + Powerlaw} model. Energy of Xe line here is $\sim$ 32.4 keV. 
The subscript bbrefl represents the bbrefl model, rdblur represents the rdblur model and dBB represents Diskbb model. 
The flux in units of 10$^{-8}$ ergs cm$^{-2}$ s$^{-1}$ is calculated in the energy band 0.8--50 keV. Errors are quoted at a 90\% confidence level.
Luminosity is in units of 10$^{38}$ erg s$^{-1}.$ assuming the distance 13 kpc for GX 17+2.} 
\label{tab2}
\centering
\begin{tabular}{ccccccccc}
\hline

\hline
%\hline
\hline
%Parameters&\multicolumn{2}{c}{A}&\multicolumn{2}{c}{B}&\multicolumn{2}{c}{C}&\multicolumn{2}{c}{D}\\
Parameters&A&B&C&\\
\hline
$kT_{in}$ (keV)\footnote{Temperature of the Diskbb model.} &1.74$_{-0.09}^{+0.10}$ &1.77$_{-0.08}^{+0.09}$ &1.75$_{-0.09}^{+0.10}$ \\ \\
$N_{dBB}$\footnote{Normalization of the Diskbb model.}& 52.19$_{-8.05}^{+9.13}$&52.57$_{-7.29}^{+7.80}$&51.38$_{-7.26}^{+8.04}$\\ \\
$R_{eff}$\footnote{Effective radius obtained by using Diskbb Normalization, spectral corrections (see text) and the inclination angle from the {\it rdblur model}}& 11.77-16.36 km&11.81-16.41 km& 11.69-16.25 km \\ \\
log$\xi$ \footnote{Ionization parameter.}&3.37 $_{-0.16}^{+0.16}$ & 3.25 $_{-0.18}^{+0.18}$&3.04 $_{-0.27}^{+0.22}$\\ \\
$kT_{bbrefl}$ (keV)\footnote{Temperature of the bbrefl model.} &2.40$_{-0.06}^{+0.07}$ &2.45$_{-0.07}^{+0.09}$ &2.37$_{-0.11}^{+0.15}$ \\ \\
f$_{refl}$\footnote{Reflection fraction.} & 0.64 $_{-0.05}^{+0.05}$& 0.62 $_{-0.05}^{+0.05}$&0.66 $_{-0.09}^{+0.09}$\\ \\
$z$\footnote{Redshift}&0&0&0\\ \\
N$_{bbrefl}$(1 $\times$10$^{-26}$)\footnote{Normalization of the bbrefl model.}& 1.85$_{-0.64}^{+0.91}$&2.00$_{-0.97}^{+1.07}$&2.42$_{-1.09}^{+1.82}$\\ \\
$\beta$$_{rdblur}$\footnote{Emissivity Index of the rdblur model.}&-3&-3&-3\\ \\
R$_{in}$(GM/c$^2$)\footnote{Inner disk radii of the rdblur model.}&8.06$_{-0.92}^{+0.94}$&7.85$_{-0.70}^{+0.95}$&7.51$_{-0.83}^{+1.15}$\\ \\
R$_{out}$(GM/c$^2$)\footnote{Outer disk radii of the rdblur model.}&1000&1000&1000\\ \\
i$^{\circ}$$_{rdblur}$\footnote{Inclination angle of the rdblur model.}&27.82$_{-3.44}^{+2.77}$ &27.61$_{-3.12}^{+2.38}$&27.91$_{-3.25}^{+2.72}$\\ \\
$\Gamma_{pl}$\footnote{Power-law index.}&2.31$_{-0.03}^{+0.03}$&2.39$_{-0.02}^{+0.02}$&2.44$_{-0.03}^{+0.03}$ \\ \\
$N_{pl}$\footnote{Normalization of the PL model.}&0.95$_{-0.10}^{+0.10}$ &0.98$_{-0.08}^{+0.08}$&0.98$_{-0.09}^{+0.09}$\\ \\
Total flux&2.42&2.29&2.05\\ \\
BB flux &0.73&0.57&0.43\\ \\
Powerlaw flux& 0.46&0.41&0.39\\ \\
$R_{sp}$\footnote{Spherization radius}&42.62 km&40.33 km &36.10 km \\ \\
$R_{BL}$\footnote{Boundary Layer radius}&71.02 km&65.98 km&57.24 km \\ \\
L$_{0.8-50 keV}$&4.87&4.61&4.13 \\ \\
$\chi^{2}$/dof&957/787&977/787&1044/787\\ \\
\hline
\end{tabular}
\end{minipage}
\end{table}
\clearpage
\begin{table}
\begin{minipage}[t]{\columnwidth}
\scriptsize
\caption{Best-fit spectral parameters for the upper, middle and lower sections of the HID which the letters A, B and C respectively
 represent using the {\it Nthcomp+ Gaussian(Fe) +  Gaussian (Xe) + Powerlaw model}. The flux in units of 10$^{-8}$ ergs cm$^{-2}$ s$^{-1}$ 
 is calculated in the energy band 0.8--50 keV. Errors are quoted at a 90\% confidence level.
Luminosity is in units of 10$^{38}$ erg s$^{-1}.$ assuming the distance 13 kpc for GX 17+2.} 
\label{tab1}
\centering
\begin{tabular}{ccccccccc}
\hline

\hline
%&&Nthcomp + Powerlaw + Gaussian&&\\
%\hline
\hline
%Parameters&\multicolumn{2}{c}{A}&\multicolumn{2}{c}{B}&\multicolumn{2}{c}{C}&\multicolumn{2}{c}{D}\\
Parameters&A&B&C&\\
\hline
$\Gamma_{Nthcomp}$\footnote{Nthcomp Power-law index.} & 2.19 $_{-0.07}^{+0.08}$& 2.29 $_{-0.07}^{+0.08}$ & 2.69 $_{-0.12}^{+0.12}$\\
$kT_{e}$ (keV)\footnote{Electron temperature (Nthcomp).} & 2.60 $_{-0.06}^{+0.07}$ & 2.63 $_{-0.07}^{+0.08}$& 2.81 $_{-0.14}^{+0.17}$  \\
$kT_{bb}$ (keV)\footnote{Seed photon temperature (Nthcomp).} & 0.96 $_{-0.03}^{+0.03}$& 0.92 $_{-0.02}^{+0.03}$ & 1.01 $_{-0.03}^{+0.03}$ \\
$N_{Nthcomp}$\footnote{Normalization of the Nthcomp model.}& 0.36 $_{-0.02}^{+0.02}$ & 0.40 $_{-0.02}^{+0.02}$& 0.33 $_{-0.01}^{+0.01}$ \\
$E_{Fe}$ keV \footnote{Line Energy of the Gaussian model for Iron line.}&6.7 &6.7 &6.7 \\ \\
$\sigma_{Fe}$\footnote{Line width of the Gaussian model for Iron line.}&0.5 &0.5&0.5\\ \\
$N_{Fe}$\footnote{Normalization of the Gaussian model for Iron line.}&0.007&0.007&0.007 \\ \\
$\Gamma_{pl}$\footnote{Power-law index.}&2.56$_{-0.01}^{+0.01}$&2.64$_{-0.01}^{+0.02}$&2.72$_{-0.02}^{+0.02}$ \\ \\
$N_{pl}$\footnote{Normalization of the PL model.}&2.29$_{-0.08}^{+0.08}$ &2.31$_{-0.07}^{+0.08}$&2.45$_{-0.08}^{+0.08}$\\ \\
Total flux & 2.46 & 2.34&2.12\\
L$_{0.8-50 keV}$&4.95&4.71&4.27 \\ \\
$\chi^{2}$/dof&910/788&994/788&928/788\\ \\
\hline
\end{tabular}
\end{minipage}
\end{table}

%%%%%%%%%%%%%%%%%%%% REFERENCES %%%%%%%%%%%%%%%%%%

\nocite{*}
\bibliography{ref}{}

\begin{thebibliography}{}
\makeatletter
\relax
\def\mn@urlcharsother{\let\do\@makeother \do\$\do\&\do\#\do\^\do\_\do\%\do\~}
\def\mn@doi{\begingroup\mn@urlcharsother \@ifnextchar [ {\mn@doi@}
  {\mn@doi@[]}}
\def\mn@doi@[#1]#2{\def\@tempa{#1}\ifx\@tempa\@empty \href
  {http://dx.doi.org/#2} {doi:#2}\else \href {http://dx.doi.org/#2} {#1}\fi
  \endgroup}
\def\mn@eprint#1#2{\mn@eprint@#1:#2::\@nil}
\def\mn@eprint@arXiv#1{\href {http://arxiv.org/abs/#1} {{\tt arXiv:#1}}}
\def\mn@eprint@dblp#1{\href {http://dblp.uni-trier.de/rec/bibtex/#1.xml}
  {dblp:#1}}
\def\mn@eprint@#1:#2:#3:#4\@nil{\def\@tempa {#1}\def\@tempb {#2}\def\@tempc
  {#3}\ifx \@tempc \@empty \let \@tempc \@tempb \let \@tempb \@tempa \fi \ifx
  \@tempb \@empty \def\@tempb {arXiv}\fi \@ifundefined
  {mn@eprint@\@tempb}{\@tempb:\@tempc}{\expandafter \expandafter \csname
  mn@eprint@\@tempb\endcsname \expandafter{\@tempc}}}

\bibitem[\protect\citeauthoryear{{Agrawal} et~al.,}{{Agrawal}
  et~al.}{2017}]{2017JApA...38...30A}
{Agrawal} P.~C.,  et~al., 2017, \mn@doi [Journal of Astrophysics and Astronomy]
  {10.1007/s12036-017-9451-z}, \href
  {https://ui.adsabs.harvard.edu/abs/2017JApA...38...30A} {38, 30}

\bibitem[\protect\citeauthoryear{{Agrawal}, {Nandi}  \& {Ramadevi}}{{Agrawal}
  et~al.}{2020}]{2020Ap&SS.365...41A}
{Agrawal} V.~K.,  {Nandi} A.,   {Ramadevi} M.~C.,  2020, \mn@doi [\apss]
  {10.1007/s10509-020-3748-0}, \href
  {https://ui.adsabs.harvard.edu/abs/2020Ap&SS.365...41A} {365, 41}

\bibitem[\protect\citeauthoryear{{Alpar} \& {Shaham}}{{Alpar} \&
  {Shaham}}{1985}]{1985Natur.316..239A}
{Alpar} M.~A.,  {Shaham} J.,  1985, \mn@doi [\nat] {10.1038/316239a0}, \href
  {https://ui.adsabs.harvard.edu/abs/1985Natur.316..239A} {316, 239}

\bibitem[\protect\citeauthoryear{{Alpar}, {Hasinger}, {Shaham}  \&
  {Yancopoulos}}{{Alpar} et~al.}{1992}]{1992A&A...257..627A}
{Alpar} M.~A.,  {Hasinger} G.,  {Shaham} J.,   {Yancopoulos} S.,  1992, \aap,
  \href {https://ui.adsabs.harvard.edu/abs/1992A&A...257..627A} {257, 627}

\bibitem[\protect\citeauthoryear{{Antia} et~al.,}{{Antia}
  et~al.}{2017}]{2017ApJS..231...10A}
{Antia} H.~M.,  et~al., 2017, \mn@doi [\apjs] {10.3847/1538-4365/aa7a0e}, \href
  {https://ui.adsabs.harvard.edu/abs/2017ApJS..231...10A} {231, 10}

\bibitem[\protect\citeauthoryear{Ballantyne}{Ballantyne}{2004}]{10.1111/j.1365-2966.2004.07767.x}
Ballantyne D.~R.,  2004, \mn@doi [Monthly Notices of the Royal Astronomical
  Society] {10.1111/j.1365-2966.2004.07767.x}, 351, 57

\bibitem[\protect\citeauthoryear{{Barret}, {Olive}, {Boirin}, {Done}, {Skinner}
   \& {Grindlay}}{{Barret} et~al.}{2000}]{2000ApJ...533..329B}
{Barret} D.,  {Olive} J.~F.,  {Boirin} L.,  {Done} C.,  {Skinner} G.~K.,
  {Grindlay} J.~E.,  2000, \mn@doi [\apj] {10.1086/308651}, \href
  {https://ui.adsabs.harvard.edu/abs/2000ApJ...533..329B} {533, 329}

\bibitem[\protect\citeauthoryear{{Beri} et~al.,}{{Beri}
  et~al.}{2019}]{2019MNRAS.482.4397B}
{Beri} A.,  et~al., 2019, \mn@doi [\mnras] {10.1093/mnras/sty2975}, \href
  {https://ui.adsabs.harvard.edu/abs/2019MNRAS.482.4397B} {482, 4397}

\bibitem[\protect\citeauthoryear{{Bhargava}, {Belloni}, {Bhattacharya}  \&
  {Misra}}{{Bhargava} et~al.}{2019}]{2019MNRAS.488..720B}
{Bhargava} Y.,  {Belloni} T.,  {Bhattacharya} D.,   {Misra} R.,  2019, \mn@doi
  [\mnras] {10.1093/mnras/stz1774}, \href
  {https://ui.adsabs.harvard.edu/abs/2019MNRAS.488..720B} {488, 720}

\bibitem[\protect\citeauthoryear{{Cackett}}{{Cackett}}{2016}]{2016ApJ...826..103C}
{Cackett} E.~M.,  2016, \mn@doi [\apj] {10.3847/0004-637X/826/2/103}, \href
  {https://ui.adsabs.harvard.edu/abs/2016ApJ...826..103C} {826, 103}

\bibitem[\protect\citeauthoryear{{Cackett} et~al.,}{{Cackett}
  et~al.}{2008}]{2008ApJ...674..415C}
{Cackett} E.~M.,  et~al., 2008, \mn@doi [\apj] {10.1086/524936}, \href
  {https://ui.adsabs.harvard.edu/abs/2008ApJ...674..415C} {674, 415}

\bibitem[\protect\citeauthoryear{{Cackett} et~al.,}{{Cackett}
  et~al.}{2009a}]{2009ApJ...690.1847C}
{Cackett} E.~M.,  et~al., 2009a, \mn@doi [\apj] {10.1088/0004-637X/690/2/1847},
  \href {https://ui.adsabs.harvard.edu/abs/2009ApJ...690.1847C} {690, 1847}

\bibitem[\protect\citeauthoryear{{Cackett}, {Altamirano}, {Patruno}, {Miller},
  {Reynolds}, {Linares}  \& {Wijnands}}{{Cackett}
  et~al.}{2009b}]{2009ApJ...694L..21C}
{Cackett} E.~M.,  {Altamirano} D.,  {Patruno} A.,  {Miller} J.~M.,  {Reynolds}
  M.,  {Linares} M.,   {Wijnands} R.,  2009b, \mn@doi [\apjl]
  {10.1088/0004-637X/694/1/L21}, \href
  {https://ui.adsabs.harvard.edu/abs/2009ApJ...694L..21C} {694, L21}

\bibitem[\protect\citeauthoryear{{Cackett} et~al.,}{{Cackett}
  et~al.}{2010}]{2010ApJ...720..205C}
{Cackett} E.~M.,  et~al., 2010, \mn@doi [\apj] {10.1088/0004-637X/720/1/205},
  \href {https://ui.adsabs.harvard.edu/abs/2010ApJ...720..205C} {720, 205}

\bibitem[\protect\citeauthoryear{{Casella}, {Belloni}  \& {Stella}}{{Casella}
  et~al.}{2006}]{2006A&A...446..579C}
{Casella} P.,  {Belloni} T.,   {Stella} L.,  2006, \mn@doi [\aap]
  {10.1051/0004-6361:20052912}, \href
  {https://ui.adsabs.harvard.edu/abs/2006A&A...446..579C} {446, 579}

\bibitem[\protect\citeauthoryear{{Church}}{{Church}}{2001}]{2001AdSpR..28..323C}
{Church} M.~J.,  2001, \mn@doi [Advances in Space Research]
  {10.1016/S0273-1177(01)00415-X}, \href
  {https://ui.adsabs.harvard.edu/abs/2001AdSpR..28..323C} {28, 323}

\bibitem[\protect\citeauthoryear{{Church} \& {Balucinska-Church}}{{Church} \&
  {Balucinska-Church}}{1995}]{1995A&A...300..441C}
{Church} M.~J.,  {Balucinska-Church} M.,  1995, \aap, \href
  {https://ui.adsabs.harvard.edu/abs/1995A&A...300..441C} {300, 441}

\bibitem[\protect\citeauthoryear{{Church} \&
  {Ba{\l}uci{\'n}ska-Church}}{{Church} \&
  {Ba{\l}uci{\'n}ska-Church}}{2004}]{2004MNRAS.348..955C}
{Church} M.~J.,  {Ba{\l}uci{\'n}ska-Church} M.,  2004, \mn@doi [\mnras]
  {10.1111/j.1365-2966.2004.07162.x}, \href
  {https://ui.adsabs.harvard.edu/abs/2004MNRAS.348..955C} {348, 955}

\bibitem[\protect\citeauthoryear{{Church}, {Gibiec}, {Ba{\l}uci{\'n}ska-Church}
   \& {Jackson}}{{Church} et~al.}{2012}]{2012A&A...546A..35C}
{Church} M.~J.,  {Gibiec} A.,  {Ba{\l}uci{\'n}ska-Church} M.,   {Jackson}
  N.~K.,  2012, \mn@doi [\aap] {10.1051/0004-6361/201218987}, \href
  {https://ui.adsabs.harvard.edu/abs/2012A&A...546A..35C} {546, A35}

\bibitem[\protect\citeauthoryear{{Davis}, {Blaes}, {Hubeny}  \&
  {Turner}}{{Davis} et~al.}{2005}]{2005ApJ...621..372D}
{Davis} S.~W.,  {Blaes} O.~M.,  {Hubeny} I.,   {Turner} N.~J.,  2005, \mn@doi
  [\apj] {10.1086/427278}, \href
  {https://ui.adsabs.harvard.edu/abs/2005ApJ...621..372D} {621, 372}

\bibitem[\protect\citeauthoryear{{Di Salvo} et~al.,}{{Di Salvo}
  et~al.}{2000}]{2000ApJ...544L.119D}
{Di Salvo} T.,  et~al., 2000, \mn@doi [\apjl] {10.1086/317309}, \href
  {https://ui.adsabs.harvard.edu/abs/2000ApJ...544L.119D} {544, L119}

\bibitem[\protect\citeauthoryear{{Di Salvo} et~al.,}{{Di Salvo}
  et~al.}{2002}]{2002A&A...386..535D}
{Di Salvo} T.,  et~al., 2002, \mn@doi [\aap] {10.1051/0004-6361:20020238},
  \href {https://ui.adsabs.harvard.edu/abs/2002A&A...386..535D} {386, 535}

\bibitem[\protect\citeauthoryear{{Dieters} \& {van der Klis}}{{Dieters} \& {van
  der Klis}}{2000}]{2000MNRAS.311..201D}
{Dieters} S.~W.,  {van der Klis} M.,  2000, \mn@doi [\mnras]
  {10.1046/j.1365-8711.2000.03050.x}, \href
  {https://ui.adsabs.harvard.edu/abs/2000MNRAS.311..201D} {311, 201}

\bibitem[\protect\citeauthoryear{{Ding}, {Zhang}, {Wang}, {Qu}  \&
  {Yan}}{{Ding} et~al.}{2011}]{2011AJ....142...34D}
{Ding} G.~Q.,  {Zhang} S.~N.,  {Wang} N.,  {Qu} J.~L.,   {Yan} S.~P.,  2011,
  \mn@doi [\aj] {10.1088/0004-6256/142/2/34}, \href
  {https://ui.adsabs.harvard.edu/abs/2011AJ....142...34D} {142, 34}

\bibitem[\protect\citeauthoryear{{Fabian}, {Rees}, {Stella}  \&
  {White}}{{Fabian} et~al.}{1989}]{1989MNRAS.238..729F}
{Fabian} A.~C.,  {Rees} M.~J.,  {Stella} L.,   {White} N.~E.,  1989, \mn@doi
  [\mnras] {10.1093/mnras/238.3.729}, \href
  {https://ui.adsabs.harvard.edu/abs/1989MNRAS.238..729F} {238, 729}

\bibitem[\protect\citeauthoryear{{Farinelli}, {Titarchuk}, {Paizis}  \&
  {Frontera}}{{Farinelli} et~al.}{2008}]{2008ApJ...680..602F}
{Farinelli} R.,  {Titarchuk} L.,  {Paizis} A.,   {Frontera} F.,  2008, \mn@doi
  [\apj] {10.1086/587162}, \href
  {https://ui.adsabs.harvard.edu/abs/2008ApJ...680..602F} {680, 602}

\bibitem[\protect\citeauthoryear{{Fender}, {Homan}  \& {Belloni}}{{Fender}
  et~al.}{2009}]{2009MNRAS.396.1370F}
{Fender} R.~P.,  {Homan} J.,   {Belloni} T.~M.,  2009, \mn@doi [\mnras]
  {10.1111/j.1365-2966.2009.14841.x}, \href
  {https://ui.adsabs.harvard.edu/abs/2009MNRAS.396.1370F} {396, 1370}

\bibitem[\protect\citeauthoryear{{Fortner}, {Lamb}  \& {Miller}}{{Fortner}
  et~al.}{1989}]{1989Natur.342..775F}
{Fortner} B.,  {Lamb} F.~K.,   {Miller} G.~S.,  1989, \mn@doi [\nat]
  {10.1038/342775a0}, \href
  {https://ui.adsabs.harvard.edu/abs/1989Natur.342..775F} {342, 775}

\bibitem[\protect\citeauthoryear{{Galloway}, {Muno}, {Hartman}, {Psaltis}  \&
  {Chakrabarty}}{{Galloway} et~al.}{2008}]{2008ApJS..179..360G}
{Galloway} D.~K.,  {Muno} M.~P.,  {Hartman} J.~M.,  {Psaltis} D.,
  {Chakrabarty} D.,  2008, \mn@doi [\apjs] {10.1086/592044}, \href
  {https://ui.adsabs.harvard.edu/abs/2008ApJS..179..360G} {179, 360}

\bibitem[\protect\citeauthoryear{{Gilfanov}, {Revnivtsev}  \&
  {Molkov}}{{Gilfanov} et~al.}{2003}]{2003A&A...410..217G}
{Gilfanov} M.,  {Revnivtsev} M.,   {Molkov} S.,  2003, \mn@doi [\aap]
  {10.1051/0004-6361:20031141}, \href
  {https://ui.adsabs.harvard.edu/abs/2003A&A...410..217G} {410, 217}

\bibitem[\protect\citeauthoryear{{Hasinger}}{{Hasinger}}{1987}]{1987A&A...186..153H}
{Hasinger} G.,  1987, \aap, \href
  {https://ui.adsabs.harvard.edu/abs/1987A&A...186..153H} {186, 153}

\bibitem[\protect\citeauthoryear{{Hasinger} \& {van der Klis}}{{Hasinger} \&
  {van der Klis}}{1989}]{1989A&A...225...79H}
{Hasinger} G.,  {van der Klis} M.,  1989, \aap, \href
  {https://ui.adsabs.harvard.edu/abs/1989A&A...225...79H} {225, 79}

\bibitem[\protect\citeauthoryear{{Hasinger}, {van der Klis}, {Ebisawa},
  {Dotani}  \& {Mitsuda}}{{Hasinger} et~al.}{1990}]{1990A&A...235..131H}
{Hasinger} G.,  {van der Klis} M.,  {Ebisawa} K.,  {Dotani} T.,   {Mitsuda} K.,
   1990, \aap, \href {https://ui.adsabs.harvard.edu/abs/1990A&A...235..131H}
  {235, 131}

\bibitem[\protect\citeauthoryear{{Homan}, {van der Klis}, {Jonker}, {Wijnands},
  {Kuulkers}, {M{\'e}ndez}  \& {Lewin}}{{Homan}
  et~al.}{2002}]{2002ApJ...568..878H}
{Homan} J.,  {van der Klis} M.,  {Jonker} P.~G.,  {Wijnands} R.,  {Kuulkers}
  E.,  {M{\'e}ndez} M.,   {Lewin} W. H.~G.,  2002, \mn@doi [\apj]
  {10.1086/339057}, \href
  {https://ui.adsabs.harvard.edu/abs/2002ApJ...568..878H} {568, 878}

\bibitem[\protect\citeauthoryear{{Homan}, {Steiner}, {Lin}, {Fridriksson},
  {Remillard}, {Miller}  \& {Ludlam}}{{Homan}
  et~al.}{2018}]{2018ApJ...853..157H}
{Homan} J.,  {Steiner} J.~F.,  {Lin} D.,  {Fridriksson} J.~K.,  {Remillard}
  R.~A.,  {Miller} J.~M.,   {Ludlam} R.~M.,  2018, \mn@doi [\apj]
  {10.3847/1538-4357/aaa439}, \href
  {https://ui.adsabs.harvard.edu/abs/2018ApJ...853..157H} {853, 157}

\bibitem[\protect\citeauthoryear{{Jithesh}, {Maqbool}, {Misra}, {T}, {Mall}  \&
  {James}}{{Jithesh} et~al.}{2019}]{2019ApJ...887..101J}
{Jithesh} V.,  {Maqbool} B.,  {Misra} R.,  {T} A.~R.,  {Mall} G.,   {James} M.,
   2019, \mn@doi [\apj] {10.3847/1538-4357/ab4f6a}, \href
  {https://ui.adsabs.harvard.edu/abs/2019ApJ...887..101J} {887, 101}

\bibitem[\protect\citeauthoryear{{Jonker}, {Wijnands}, {van der Klis},
  {Psaltis}, {Kuulkers}  \& {Lamb}}{{Jonker}
  et~al.}{1998}]{1998ApJ...499L.191J}
{Jonker} P.~G.,  {Wijnands} R.,  {van der Klis} M.,  {Psaltis} D.,  {Kuulkers}
  E.,   {Lamb} F.~K.,  1998, \mn@doi [\apjl] {10.1086/311372}, \href
  {https://ui.adsabs.harvard.edu/abs/1998ApJ...499L.191J} {499, L191}

\bibitem[\protect\citeauthoryear{{Kara} et~al.,}{{Kara}
  et~al.}{2019}]{2019Natur.565..198K}
{Kara} E.,  et~al., 2019, \mn@doi [\nat] {10.1038/s41586-018-0803-x}, \href
  {https://ui.adsabs.harvard.edu/abs/2019Natur.565..198K} {565, 198}

\bibitem[\protect\citeauthoryear{{Kotov}, {Churazov}  \& {Gilfanov}}{{Kotov}
  et~al.}{2001}]{2001MNRAS.327..799K}
{Kotov} O.,  {Churazov} E.,   {Gilfanov} M.,  2001, \mn@doi [\mnras]
  {10.1046/j.1365-8711.2001.04769.x}, \href
  {https://ui.adsabs.harvard.edu/abs/2001MNRAS.327..799K} {327, 799}

\bibitem[\protect\citeauthoryear{{Kubota}, {Tanaka}, {Makishima}, {Ueda},
  {Dotani}, {Inoue}  \& {Yamaoka}}{{Kubota} et~al.}{1998}]{1998PASJ...50..667K}
{Kubota} A.,  {Tanaka} Y.,  {Makishima} K.,  {Ueda} Y.,  {Dotani} T.,  {Inoue}
  H.,   {Yamaoka} K.,  1998, \mn@doi [\pasj] {10.1093/pasj/50.6.667}, \href
  {https://ui.adsabs.harvard.edu/abs/1998PASJ...50..667K} {50, 667}

\bibitem[\protect\citeauthoryear{{Kubota}, {Makishima}  \& {Ebisawa}}{{Kubota}
  et~al.}{2001}]{2001ApJ...560L.147K}
{Kubota} A.,  {Makishima} K.,   {Ebisawa} K.,  2001, \mn@doi [\apjl]
  {10.1086/324377}, \href
  {https://ui.adsabs.harvard.edu/abs/2001ApJ...560L.147K} {560, L147}

\bibitem[\protect\citeauthoryear{{Kuulkers} \& {van der Klis}}{{Kuulkers} \&
  {van der Klis}}{1995}]{1995A&A...303..801K}
{Kuulkers} E.,  {van der Klis} M.,  1995, \aap, \href
  {https://ui.adsabs.harvard.edu/abs/1995A&A...303..801K} {303, 801}

\bibitem[\protect\citeauthoryear{{Kuulkers}, {van der Klis}, {Oosterbroek},
  {van Paradijs}  \& {Lewin}}{{Kuulkers} et~al.}{1997}]{1997MNRAS.287..495K}
{Kuulkers} E.,  {van der Klis} M.,  {Oosterbroek} T.,  {van Paradijs} J.,
  {Lewin} W.~H.~G.,  1997, \mn@doi [\mnras] {10.1093/mnras/287.3.495}, \href
  {https://ui.adsabs.harvard.edu/abs/1997MNRAS.287..495K} {287, 495}

\bibitem[\protect\citeauthoryear{{Kuulkers}, {den Hartog}, {in't Zand},
  {Verbunt}, {Harris}  \& {Cocchi}}{{Kuulkers}
  et~al.}{2003}]{2003A&A...399..663K}
{Kuulkers} E.,  {den Hartog} P.~R.,  {in't Zand} J.~J.~M.,  {Verbunt} F.~W.~M.,
   {Harris} W.~E.,   {Cocchi} M.,  2003, \mn@doi [\aap]
  {10.1051/0004-6361:20021781}, \href
  {https://ui.adsabs.harvard.edu/abs/2003A&A...399..663K} {399, 663}

\bibitem[\protect\citeauthoryear{{Lamb}}{{Lamb}}{1989}]{1989ESASP.296..215L}
{Lamb} F.~K.,  1989, in {Hunt} J.,  {Battrick} B.,  eds,  ESA Special
  Publication Vol. 1, Two Topics in X-Ray Astronomy, Volume 1: X Ray Binaries.
  Volume 2: AGN and the X Ray Background. p.~215

\bibitem[\protect\citeauthoryear{{Lamb}, {Shibazaki}, {Alpar}  \&
  {Shaham}}{{Lamb} et~al.}{1985}]{1985Natur.317..681L}
{Lamb} F.~K.,  {Shibazaki} N.,  {Alpar} M.~A.,   {Shaham} J.,  1985, \mn@doi
  [\nat] {10.1038/317681a0}, \href
  {https://ui.adsabs.harvard.edu/abs/1985Natur.317..681L} {317, 681}

\bibitem[\protect\citeauthoryear{{Lei} et~al.,}{{Lei}
  et~al.}{2008}]{2008ApJ...677..461L}
{Lei} Y.~J.,  et~al., 2008, \mn@doi [\apj] {10.1086/533423}, \href
  {https://ui.adsabs.harvard.edu/abs/2008ApJ...677..461L} {677, 461}

\bibitem[\protect\citeauthoryear{{Lin}, {Remillard}  \& {Homan}}{{Lin}
  et~al.}{2007}]{2007ApJ...667.1073L}
{Lin} D.,  {Remillard} R.~A.,   {Homan} J.,  2007, \mn@doi [\apj]
  {10.1086/521181}, \href
  {https://ui.adsabs.harvard.edu/abs/2007ApJ...667.1073L} {667, 1073}

\bibitem[\protect\citeauthoryear{{Lin}, {Remillard}  \& {Homan}}{{Lin}
  et~al.}{2009}]{2009ApJ...696.1257L}
{Lin} D.,  {Remillard} R.~A.,   {Homan} J.,  2009, \mn@doi [\apj]
  {10.1088/0004-637X/696/2/1257}, \href
  {https://ui.adsabs.harvard.edu/abs/2009ApJ...696.1257L} {696, 1257}

\bibitem[\protect\citeauthoryear{{Ludlam} et~al.,}{{Ludlam}
  et~al.}{2017a}]{2017ApJ...836..140L}
{Ludlam} R.~M.,  et~al., 2017a, \mn@doi [\apj] {10.3847/1538-4357/836/1/140},
  \href {https://ui.adsabs.harvard.edu/abs/2017ApJ...836..140L} {836, 140}

\bibitem[\protect\citeauthoryear{{Ludlam}, {Miller}, {Degenaar}, {Sanna},
  {Cackett}, {Altamirano}  \& {King}}{{Ludlam}
  et~al.}{2017b}]{2017ApJ...847..135L}
{Ludlam} R.~M.,  {Miller} J.~M.,  {Degenaar} N.,  {Sanna} A.,  {Cackett} E.~M.,
   {Altamirano} D.,   {King} A.~L.,  2017b, \mn@doi [\apj]
  {10.3847/1538-4357/aa8b1b}, \href
  {https://ui.adsabs.harvard.edu/abs/2017ApJ...847..135L} {847, 135}

\bibitem[\protect\citeauthoryear{{Manmoto}, {Mineshige}  \&
  {Kusunose}}{{Manmoto} et~al.}{1997}]{1997ApJ...489..791M}
{Manmoto} T.,  {Mineshige} S.,   {Kusunose} M.,  1997, \mn@doi [\apj]
  {10.1086/304817}, \href
  {https://ui.adsabs.harvard.edu/abs/1997ApJ...489..791M} {489, 791}

\bibitem[\protect\citeauthoryear{McKinney, Tchekhovskoy  \& Blandford}{McKinney
  et~al.}{2012}]{10.1111/j.1365-2966.2012.21074.x}
McKinney J.~C.,  Tchekhovskoy A.,   Blandford R.~D.,  2012, \mn@doi [Monthly
  Notices of the Royal Astronomical Society]
  {10.1111/j.1365-2966.2012.21074.x}, 423, 3083

\bibitem[\protect\citeauthoryear{{Migliari} et~al.,}{{Migliari}
  et~al.}{2007}]{2007ApJ...671..706M}
{Migliari} S.,  et~al., 2007, \mn@doi [\apj] {10.1086/522516}, \href
  {https://ui.adsabs.harvard.edu/abs/2007ApJ...671..706M} {671, 706}

\bibitem[\protect\citeauthoryear{{Migliari}, {Miller-Jones}  \&
  {Russell}}{{Migliari} et~al.}{2011}]{2011MNRAS.415.2407M}
{Migliari} S.,  {Miller-Jones} J.~C.~A.,   {Russell} D.~M.,  2011, \mn@doi
  [\mnras] {10.1111/j.1365-2966.2011.18868.x}, \href
  {https://ui.adsabs.harvard.edu/abs/2011MNRAS.415.2407M} {415, 2407}

\bibitem[\protect\citeauthoryear{{Mitsuda} et~al.,}{{Mitsuda}
  et~al.}{1984}]{1984PASJ...36..741M}
{Mitsuda} K.,  et~al., 1984, \pasj, \href
  {https://ui.adsabs.harvard.edu/abs/1984PASJ...36..741M} {36, 741}

\bibitem[\protect\citeauthoryear{{Miyamoto}, {Kimura}, {Kitamoto}, {Dotani}  \&
  {Ebisawa}}{{Miyamoto} et~al.}{1991}]{1991ApJ...383..784M}
{Miyamoto} S.,  {Kimura} K.,  {Kitamoto} S.,  {Dotani} T.,   {Ebisawa} K.,
  1991, \mn@doi [\apj] {10.1086/170837}, \href
  {https://ui.adsabs.harvard.edu/abs/1991ApJ...383..784M} {383, 784}

\bibitem[\protect\citeauthoryear{{Mondal}, {Dewangan}, {Pahari}, {Misra},
  {Kembhavi}  \& {Raychaudhuri}}{{Mondal} et~al.}{2016}]{2016MNRAS.461.1917M}
{Mondal} A.~S.,  {Dewangan} G.~C.,  {Pahari} M.,  {Misra} R.,  {Kembhavi}
  A.~K.,   {Raychaudhuri} B.,  2016, \mn@doi [\mnras] {10.1093/mnras/stw1464},
  \href {https://ui.adsabs.harvard.edu/abs/2016MNRAS.461.1917M} {461, 1917}

\bibitem[\protect\citeauthoryear{{Motta} \& {Fender}}{{Motta} \&
  {Fender}}{2019}]{2019MNRAS.483.3686M}
{Motta} S.~E.,  {Fender} R.~P.,  2019, \mn@doi [\mnras]
  {10.1093/mnras/sty3331}, \href
  {https://ui.adsabs.harvard.edu/abs/2019MNRAS.483.3686M} {483, 3686}

\bibitem[\protect\citeauthoryear{{Pen}, {Matzner}  \& {Wong}}{{Pen}
  et~al.}{2003}]{2003ApJ...596L.207P}
{Pen} U.-L.,  {Matzner} C.~D.,   {Wong} S.,  2003, \mn@doi [\apjl]
  {10.1086/379339}, \href
  {https://ui.adsabs.harvard.edu/abs/2003ApJ...596L.207P} {596, L207}

\bibitem[\protect\citeauthoryear{{Penninx}, {Lewin}, {Mitsuda}, {van der Klis},
  {van Paradijs}  \& {Zilstra}}{{Penninx} et~al.}{1990}]{1990MNRAS.243..114P}
{Penninx} W.,  {Lewin} W.~H.~G.,  {Mitsuda} K.,  {van der Klis} M.,  {van
  Paradijs} J.,   {Zilstra} A.~A.,  1990, \mn@doi [\mnras]
  {10.1093/mnras/243.1.114}, \href
  {https://ui.adsabs.harvard.edu/abs/1990MNRAS.243..114P} {243, 114}

\bibitem[\protect\citeauthoryear{{Popham} \& {Sunyaev}}{{Popham} \&
  {Sunyaev}}{2001}]{2001ApJ...547..355P}
{Popham} R.,  {Sunyaev} R.,  2001, \mn@doi [\apj] {10.1086/318336}, \href
  {https://ui.adsabs.harvard.edu/abs/2001ApJ...547..355P} {547, 355}

\bibitem[\protect\citeauthoryear{{Priedhorsky}, {Hasinger}, {Lewin},
  {Middleditch}, {Parmar}, {Stella}  \& {White}}{{Priedhorsky}
  et~al.}{1986}]{1986ApJ...306L..91P}
{Priedhorsky} W.,  {Hasinger} G.,  {Lewin} W.~H.~G.,  {Middleditch} J.,
  {Parmar} A.,  {Stella} L.,   {White} N.,  1986, \mn@doi [\apjl]
  {10.1086/184712}, \href
  {https://ui.adsabs.harvard.edu/abs/1986ApJ...306L..91P} {306, L91}

\bibitem[\protect\citeauthoryear{{Pringle}}{{Pringle}}{1981}]{1981ARA&A..19..137P}
{Pringle} J.~E.,  1981, \mn@doi [\araa] {10.1146/annurev.aa.19.090181.001033},
  \href {https://ui.adsabs.harvard.edu/abs/1981ARA&A..19..137P} {19, 137}

\bibitem[\protect\citeauthoryear{{Reig} \& {Kylafis}}{{Reig} \&
  {Kylafis}}{2016}]{2016A&A...591A..24R}
{Reig} P.,  {Kylafis} N.,  2016, \mn@doi [\aap] {10.1051/0004-6361/201628294},
  \href {https://ui.adsabs.harvard.edu/abs/2016A&A...591A..24R} {591, A24}

\bibitem[\protect\citeauthoryear{{Revnivtsev} \& {Gilfanov}}{{Revnivtsev} \&
  {Gilfanov}}{2006}]{2006A&A...453..253R}
{Revnivtsev} M.~G.,  {Gilfanov} M.~R.,  2006, \mn@doi [\aap]
  {10.1051/0004-6361:20053964}, \href
  {https://ui.adsabs.harvard.edu/abs/2006A&A...453..253R} {453, 253}

\bibitem[\protect\citeauthoryear{{Reynolds} \& {Miller}}{{Reynolds} \&
  {Miller}}{2013}]{2013ApJ...769...16R}
{Reynolds} M.~T.,  {Miller} J.~M.,  2013, \mn@doi [\apj]
  {10.1088/0004-637X/769/1/16}, \href
  {https://ui.adsabs.harvard.edu/abs/2013ApJ...769...16R} {769, 16}

\bibitem[\protect\citeauthoryear{{Sanna} et~al.,}{{Sanna}
  et~al.}{2010}]{2010MNRAS.408..622S}
{Sanna} A.,  et~al., 2010, \mn@doi [\mnras] {10.1111/j.1365-2966.2010.17145.x},
  \href {https://ui.adsabs.harvard.edu/abs/2010MNRAS.408..622S} {408, 622}

\bibitem[\protect\citeauthoryear{{Shakura} \& {Sunyaev}}{{Shakura} \&
  {Sunyaev}}{1988}]{1988AdSpR...8..135S}
{Shakura} N.~I.,  {Sunyaev} R.~A.,  1988, \mn@doi [Advances in Space Research]
  {10.1016/0273-1177(88)90396-1}, \href
  {https://ui.adsabs.harvard.edu/abs/1988AdSpR...8..135S} {8, 135}

\bibitem[\protect\citeauthoryear{{Sharma}, {Beri}, {Sanna}  \&
  {Dutta}}{{Sharma} et~al.}{2020}]{2020MNRAS.492.4361S}
{Sharma} R.,  {Beri} A.,  {Sanna} A.,   {Dutta} A.,  2020, \mn@doi [\mnras]
  {10.1093/mnras/staa109}, \href
  {https://ui.adsabs.harvard.edu/abs/2020MNRAS.492.4361S} {492, 4361}

\bibitem[\protect\citeauthoryear{{Shimura} \& {Takahara}}{{Shimura} \&
  {Takahara}}{1995}]{1995ApJ...445..780S}
{Shimura} T.,  {Takahara} F.,  1995, \mn@doi [\apj] {10.1086/175740}, \href
  {https://ui.adsabs.harvard.edu/abs/1995ApJ...445..780S} {445, 780}

\bibitem[\protect\citeauthoryear{{Singh} et~al.,}{{Singh}
  et~al.}{2017}]{2017JApA...38...29S}
{Singh} K.~P.,  et~al., 2017, \mn@doi [Journal of Astrophysics and Astronomy]
  {10.1007/s12036-017-9448-7}, \href
  {https://ui.adsabs.harvard.edu/abs/2017JApA...38...29S} {38, 29}

\bibitem[\protect\citeauthoryear{{Sridhar}, {Bhattacharyya}, {Chandra}  \&
  {Antia}}{{Sridhar} et~al.}{2019}]{2019MNRAS.487.4221S}
{Sridhar} N.,  {Bhattacharyya} S.,  {Chandra} S.,   {Antia} H.~M.,  2019,
  \mn@doi [\mnras] {10.1093/mnras/stz1476}, \href
  {https://ui.adsabs.harvard.edu/abs/2019MNRAS.487.4221S} {487, 4221}

\bibitem[\protect\citeauthoryear{{Sriram}, {Agrawal}, {Pendharkar}  \&
  {Rao}}{{Sriram} et~al.}{2007}]{2007ApJ...661.1055S}
{Sriram} K.,  {Agrawal} V.~K.,  {Pendharkar} J.~K.,   {Rao} A.~R.,  2007,
  \mn@doi [\apj] {10.1086/516771}, \href
  {https://ui.adsabs.harvard.edu/abs/2007ApJ...661.1055S} {661, 1055}

\bibitem[\protect\citeauthoryear{{Sriram}, {Choi}  \& {Rao}}{{Sriram}
  et~al.}{2011a}]{2011A&A...525A.146S}
{Sriram} K.,  {Choi} C.~S.,   {Rao} A.~R.,  2011a, \mn@doi [\aap]
  {10.1051/0004-6361/201015684}, \href
  {https://ui.adsabs.harvard.edu/abs/2011A&A...525A.146S} {525, A146}

\bibitem[\protect\citeauthoryear{{Sriram}, {Rao}  \& {Choi}}{{Sriram}
  et~al.}{2011b}]{2011ApJ...743L..31S}
{Sriram} K.,  {Rao} A.~R.,   {Choi} C.~S.,  2011b, \mn@doi [\apjl]
  {10.1088/2041-8205/743/2/L31}, \href
  {https://ui.adsabs.harvard.edu/abs/2011ApJ...743L..31S} {743, L31}

\bibitem[\protect\citeauthoryear{{Sriram}, {Rao}  \& {Choi}}{{Sriram}
  et~al.}{2012}]{2012A&A...541A...6S}
{Sriram} K.,  {Rao} A.~R.,   {Choi} C.~S.,  2012, \mn@doi [\aap]
  {10.1051/0004-6361/201218799}, \href
  {https://ui.adsabs.harvard.edu/abs/2012A&A...541A...6S} {541, A6}

\bibitem[\protect\citeauthoryear{Sriram, Malu  \& Choi}{Sriram
  et~al.}{2019}]{Sriram_2019}
Sriram K.,  Malu S.,   Choi C.~S.,  2019, \mn@doi [The Astrophysical Journal
  Supplement Series] {10.3847/1538-4365/ab30e1}, 244, 5

\bibitem[\protect\citeauthoryear{{Stella} \& {Vietri}}{{Stella} \&
  {Vietri}}{1998}]{1998ApJ...492L..59S}
{Stella} L.,  {Vietri} M.,  1998, \mn@doi [\apjl] {10.1086/311075}, \href
  {https://ui.adsabs.harvard.edu/abs/1998ApJ...492L..59S} {492, L59}

\bibitem[\protect\citeauthoryear{{Stella}, {Vietri}  \& {Morsink}}{{Stella}
  et~al.}{1999}]{1999ApJ...524L..63S}
{Stella} L.,  {Vietri} M.,   {Morsink} S.~M.,  1999, \mn@doi [\apjl]
  {10.1086/312291}, \href
  {https://ui.adsabs.harvard.edu/abs/1999ApJ...524L..63S} {524, L63}

\bibitem[\protect\citeauthoryear{{Titarchuk}, {Bradshaw}, {Geldzahler}  \&
  {Fomalont}}{{Titarchuk} et~al.}{2001}]{2001ApJ...555L..45T}
{Titarchuk} L.~G.,  {Bradshaw} C.~F.,  {Geldzahler} B.~J.,   {Fomalont} E.~B.,
  2001, \mn@doi [\apjl] {10.1086/323160}, \href
  {https://ui.adsabs.harvard.edu/abs/2001ApJ...555L..45T} {555, L45}

\bibitem[\protect\citeauthoryear{{Vaughan}, {van der Klis}, {Lewin}, {van
  Paradijs}, {Mitsuda}  \& {Dotani}}{{Vaughan}
  et~al.}{1999}]{1999A&A...343..197V}
{Vaughan} B.~A.,  {van der Klis} M.,  {Lewin} W.~H.~G.,  {van Paradijs} J.,
  {Mitsuda} K.,   {Dotani} T.,  1999, \aap, \href
  {https://ui.adsabs.harvard.edu/abs/1999A&A...343..197V} {343, 197}

\bibitem[\protect\citeauthoryear{{Vrtilek}, {Raymond}, {Garcia}, {Verbunt},
  {Hasinger}  \& {Kurster}}{{Vrtilek} et~al.}{1990}]{1990A&A...235..162V}
{Vrtilek} S.~D.,  {Raymond} J.~C.,  {Garcia} M.~R.,  {Verbunt} F.,  {Hasinger}
  G.,   {Kurster} M.,  1990, \aap, \href
  {https://ui.adsabs.harvard.edu/abs/1990A&A...235..162V} {235, 162}

\bibitem[\protect\citeauthoryear{{White}, {Stella}  \& {Parmar}}{{White}
  et~al.}{1988}]{1988ApJ...324..363W}
{White} N.~E.,  {Stella} L.,   {Parmar} A.~N.,  1988, \mn@doi [\apj]
  {10.1086/165901}, \href
  {https://ui.adsabs.harvard.edu/abs/1988ApJ...324..363W} {324, 363}

\bibitem[\protect\citeauthoryear{{Wijnands} et~al.,}{{Wijnands}
  et~al.}{1997}]{1997ApJ...490L.157W}
{Wijnands} R.,  et~al., 1997, \mn@doi [\apjl] {10.1086/311039}, \href
  {https://ui.adsabs.harvard.edu/abs/1997ApJ...490L.157W} {490, L157}

\bibitem[\protect\citeauthoryear{{Wijnands} et~al.,}{{Wijnands}
  et~al.}{1998a}]{1998ApJ...493L..87W}
{Wijnands} R.,  et~al., 1998a, \mn@doi [\apjl] {10.1086/311138}, \href
  {https://ui.adsabs.harvard.edu/abs/1998ApJ...493L..87W} {493, L87}

\bibitem[\protect\citeauthoryear{{Wijnands}, {van der Klis}, {M{\'e}ndez}, {van
  Paradijs}, {Lewin}, {Lamb}, {Vaughan}  \& {Kuulkers}}{{Wijnands}
  et~al.}{1998b}]{1998ApJ...495L..39W}
{Wijnands} R.,  {van der Klis} M.,  {M{\'e}ndez} M.,  {van Paradijs} J.,
  {Lewin} W. H.~G.,  {Lamb} F.~K.,  {Vaughan} B.,   {Kuulkers} E.,  1998b,
  \mn@doi [\apjl] {10.1086/311217}, \href
  {https://ui.adsabs.harvard.edu/abs/1998ApJ...495L..39W} {495, L39}

\bibitem[\protect\citeauthoryear{{Wijnands}, {van der Klis}  \&
  {Rijkhorst}}{{Wijnands} et~al.}{1999}]{1999ApJ...512L..39W}
{Wijnands} R.,  {van der Klis} M.,   {Rijkhorst} E.-J.,  1999, \mn@doi [\apjl]
  {10.1086/311872}, \href
  {https://ui.adsabs.harvard.edu/abs/1999ApJ...512L..39W} {512, L39}

\bibitem[\protect\citeauthoryear{{Wilms}, {Allen}  \& {McCray}}{{Wilms}
  et~al.}{2000}]{2000ApJ...542..914W}
{Wilms} J.,  {Allen} A.,   {McCray} R.,  2000, \mn@doi [\apj] {10.1086/317016},
  \href {https://ui.adsabs.harvard.edu/abs/2000ApJ...542..914W} {542, 914}

\bibitem[\protect\citeauthoryear{{Yadav} et~al.,}{{Yadav}
  et~al.}{2016}]{2016SPIE.9905E..1DY}
{Yadav} J.~S.,  et~al., 2016, {Large Area X-ray Proportional Counter (LAXPC)
  instrument onboard ASTROSAT}.
p. 99051D, \mn@doi{10.1117/12.2231857}

\bibitem[\protect\citeauthoryear{{Zdziarski}, {Johnson}  \&
  {Magdziarz}}{{Zdziarski} et~al.}{1996}]{1996MNRAS.283..193Z}
{Zdziarski} A.~A.,  {Johnson} W.~N.,   {Magdziarz} P.,  1996, \mn@doi [\mnras]
  {10.1093/mnras/283.1.193}, \href
  {https://ui.adsabs.harvard.edu/abs/1996MNRAS.283..193Z} {283, 193}

\bibitem[\protect\citeauthoryear{{Zhang}, {Strohmayer}  \& {Swank}}{{Zhang}
  et~al.}{1998}]{1998ApJ...500L.167Z}
{Zhang} W.,  {Strohmayer} T.~E.,   {Swank} J.~H.,  1998, \mn@doi [\apjl]
  {10.1086/311423}, \href
  {https://ui.adsabs.harvard.edu/abs/1998ApJ...500L.167Z} {500, L167}

\bibitem[\protect\citeauthoryear{{{\.Z}ycki}, {Done}  \& {Smith}}{{{\.Z}ycki}
  et~al.}{1999}]{1999MNRAS.309..561Z}
{{\.Z}ycki} P.~T.,  {Done} C.,   {Smith} D.~A.,  1999, \mn@doi [\mnras]
  {10.1046/j.1365-8711.1999.02885.x}, \href
  {https://ui.adsabs.harvard.edu/abs/1999MNRAS.309..561Z} {309, 561}

\bibitem[\protect\citeauthoryear{{van der Klis}}{{van der
  Klis}}{1989}]{1989ARA&A..27..517V}
{van der Klis} M.,  1989, \mn@doi [\araa]
  {10.1146/annurev.aa.27.090189.002505}, \href
  {https://ui.adsabs.harvard.edu/abs/1989ARA&A..27..517V} {27, 517}

\bibitem[\protect\citeauthoryear{{van der Klis}}{{van der
  Klis}}{2006}]{2006csxs.book...39V}
{van der Klis} M.,  2006, {Rapid X-ray Variability}.
pp 39--112

\makeatother
\end{thebibliography}
 \bibliographystyle{mnras}

%%%%%%%%%%%%%%%%%%%%%%%%%%%%%%%%%%%%%%%%%%%%%%%%%%

%%%%%%%%%%%%%%%%% APPENDICES %%%%%%%%%%%%%%%%%%%%%
\bsp	% typesetting comment
\label{lastpage}
\end{document}